\documentclass[journal]{IEEEtran}
\usepackage{amsmath,amsfonts,amssymb,amsthm}
\usepackage{algorithmic}
\usepackage{algorithm}
\usepackage{array}
\usepackage{textcomp}
\usepackage{stfloats}
\usepackage{url}
\usepackage{verbatim,makecell}
\usepackage{graphicx}
\usepackage{cite}
\usepackage{bm}
\usepackage{xcolor} 
\usepackage{subfigure}
\usepackage{amsthm}
\usepackage{upgreek} 

\newtheorem{proposition}{Proposition}

\newtheorem{remark}{Remark}

\begin{document}

\title{Baseband-Efficient  WMMSE Precoding: From a Signal Weighting Cost Perspective}

\author{Shuai Gao, Fan Xu, Mian Li, Xinzhi Ning, Lei Qiu, Boyu Ning and Qingjiang Shi
        % <-this % stops a space
%\emph{ (Corresponding author: Qingjiang Shi and Fan Xu.)}}
\thanks{S. Gao, F. Xu and L. Qiu are with the School of Electronic and Information Engineering, Tongji University, Shanghai, China (emails: gaoshuai20230901@163.com, xxiaof999@tongji.edu.cn, qiulei@tongji.edu.cn).}% <-this % stops a space
\thanks{Mian Li is with the School of Science and Engineering, The Chinese University of Hong Kong, Shenzhen, and also with Shenzhen Research Institute of Big Data, Shenzhen 518172, China (e-mail: mianli1@link.cuhk.edu.cn).}
\thanks{B. Ning is with the Department of Computer Science, KTH Royal Institute of Technology, 100 44 Stockholm, Sweden (e-mail: boydning@outlook.com).}
\thanks{X. Ning and Q. Shi are with the School of Computer Science and Technology, Tongji University, Shanghai 201804, China; Q. Shi is also with the Shenzhen Research Institute of Big Data, Shenzhen 518172, China (email:  ningxinzhi@tongji.edu.cn, shiqj@tongji.edu.cn).}
}

\maketitle
\begin{abstract}
For downlink transmission in massive multi-user multiple-input multiple-output (MU-MIMO) systems, conventional precoding research heavily focuses on reducing the computational complexity of precoding matrix design, while largely overlooking another critical bottleneck: the substantial signal weighting cost incurred by repeatedly applying the precoder to high-speed data streams. 
To address both challenges simultaneously, this paper proposes a novel sparse precoding framework tailored for fully-digital architectures. \textcolor{black}{Within this framework, from the sum-rate maximization perspective, we design two sparse precoding architectures: a common-support row-sparse architecture and a user-specific row-sparse architecture, so as to reduce the number of multiplication operations required in baseband signal weighting while largely preserving the achievable sum-rate.} For the formulated mixed-integer non-linear programming (MINLP) problem, \textcolor{black}{we rigorously prove that} the optimal precoder under both sparse architectures strictly resides in a specific low-dimensional subspace determined by the channel matrices, thereby  reducing the dimensionality of the optimization variables. Based on this  insight, an  alternating optimization algorithm is developed within the  weighted minimum mean square error (WMMSE) framework to jointly optimize sparse beam selection and low-dimensional precoding coefficients. The combinatorial beam selection problem is handled using an efficient penalty-based majorize-minimization (MM) method, yielding a low-complexity closed-form solution. \textcolor{black}{Simulation results demonstrate that the proposed schemes achieve sum-rate performance close to that of full-dimensional WMMSE precoding without sparsity constraints, while substantially reducing the overall signal-weighting cost.}
\end{abstract}


\section{Introduction} \IEEEPARstart{M}{assive} multiple-input multiple-output (MIMO) is a key enabling technology for 5G and future wireless communication systems\cite{dreifuerst2023massive,de2022overview,wang2023extremely,wang2024tutorial}. By deploying large-scale antenna arrays at the base station (BS), MIMO systems can fully exploit spatial resources, resulting in substantial improvements in spectral and energy efficiency\cite{ngo2013energy,lu2014overview}. 

Downlink precoding plays a crucial role in realizing the full potential of massive MIMO systems. In multi-user MIMO (MU-MIMO) scenarios, the BS exploits its high spatial resolution to serve multiple user terminals simultaneously. The main objective of precoding is to jointly enhance the desired signal power and suppress multi-user interference, thereby improving the overall system throughput. Sum-rate maximization, or more generally weighted sum-rate (WSR) maximization, has become a fundamental design criterion for MU-MIMO systems. Among existing approaches, the weighted minimum mean square error (WMMSE) algorithm provides a well-established framework for WSR maximization \cite{christensen2008weighted,shi2011iteratively}. \textcolor{black}{Through iterative optimization of the transmit precoders, receive filters, and weight matrices, WMMSE reliably converges to a stationary point and generally provides strong WSR performance in practical scenarios.}

Despite its strong performance, the WMMSE algorithm suffers from high computational complexity in massive MIMO systems, where the high antenna dimensionality makes the precoder update particularly costly. To address this, extensive research has focused on reducing the precoding computation complexity through various low-complexity algorithms \textcolor{black}{\cite{zarei2013low,zhang2022deep,shi2023robust,hu2020iterative,zhao2023rethinking}}. However, even when these schemes successfully lower the cost of computing the precoder itself, they often overlook a far more severe bottleneck in practical baseband deployments: the signal weighting cost.

In practical systems, a staggering execution asymmetry exists. The precoding matrix is computed only once per channel coherence interval, yet it must be repeatedly applied to thousands of high-speed data symbols \cite{dahlman20205g,gonultacs2021hardware,tong2023low}. 
As illustrated by the pie chart in Fig. 1 for a typical scenario with a 5 ms Sounding Reference Signal (SRS) period and an 8-Resource Block (RB) scheduling scope, this high-frequency matrix–vector multiplication overwhelmingly dominates the baseband processing workload. Specifically, the execution ratio of signal weighting to precoder updates reaches a striking 11,520:1 (a detailed numerical analysis is provided in Section II-B). Consequently, applying a dense precoding matrix to such massive data streams imposes a substantial computational burden on the baseband processor. This hidden cost significantly inflates the baseband processing load, thereby severely constraining the maximum supportable baseband specifications.

An effective way to reduce the signal weighting cost is to design sparse precoding matrices \cite{gonultacs2021hardware}\cite{tong2023low}, where a large fraction of precoder entries are forced to zero so as to reduce the number of multiplications required in real-time signal weighting. One major line of work arises from hybrid analog-digital architectures, where the limited number of RF chains naturally induces sparsity in the digital precoder to match low-dimensional signal processing units \cite{el2014spatially}\cite{alkhateeb2014mimo}. In such settings, precoding is often formulated as a matrix approximation or factorization problem to reconstruct the fully-digital precoder using a sparse combination of basis vectors \cite{yu2016alternating}. Another line of work focuses on sparse beamforming in the antenna domain, typically for objectives such as antenna selection, interference management, or backhaul overhead reduction, rather than signal weighting cost in fully-digital systems. For example, \cite{zhao2018joint} and \cite{huang2023sparse} impose $l_0$- and $l_1$-norm constraints, respectively, while the S-WMMSE method in \cite{hong2013joint} promotes group sparsity via mixed $l_2/l_1$ regularization. More recently, angle-domain sparse design has also attracted considerable attention, especially in mmWave and massive MIMO systems, where channel energy is naturally concentrated on a small number of dominant beams \cite{guo2018joint,hu2021joint,sun2022joint,cheng2020low,zhao2017angle}. Although these works demonstrate the potential of sparse precoding, they are primarily driven by RF hardware constraints, rather than by the signal weighting cost in fully-digital baseband processing.

\textcolor{black}{Hardware-aware beamspace precoding \cite{gonultacs2021hardware} is based on MSE-oriented sparse approximation, while low-complexity angle-domain precoding \cite{tong2023low} combines channel-energy-based beam selection with MMSE precoding. These methods do not directly formulate sparse precoding under the sum-rate maximization criterion or establish the corresponding optimal low-dimensional subspace structure. Meanwhile, conventional sparsity-constrained WMMSE precoding, such as S-WMMSE \cite{hong2013joint}, operates in the full-dimensional precoder space and incurs a per-iteration complexity of $\mathcal{O}(M^3)$.}

\textcolor{black}{Motivated by these observations, this paper proposes a novel sparse precoding framework for fully-digital architectures from the sum-rate maximization perspective.} \textcolor{black}{The proposed framework reduces the precoder-design complexity relative to conventional sparsity-constrained WMMSE precoding, while reducing the signal-weighting cost relative to dense precoding.} The main contributions of this study are summarized as follows:

\begin{figure}[t] 
\centering
\hspace*{-0.04\textwidth} % 可调节偏移量
\includegraphics[width=0.3\textwidth]{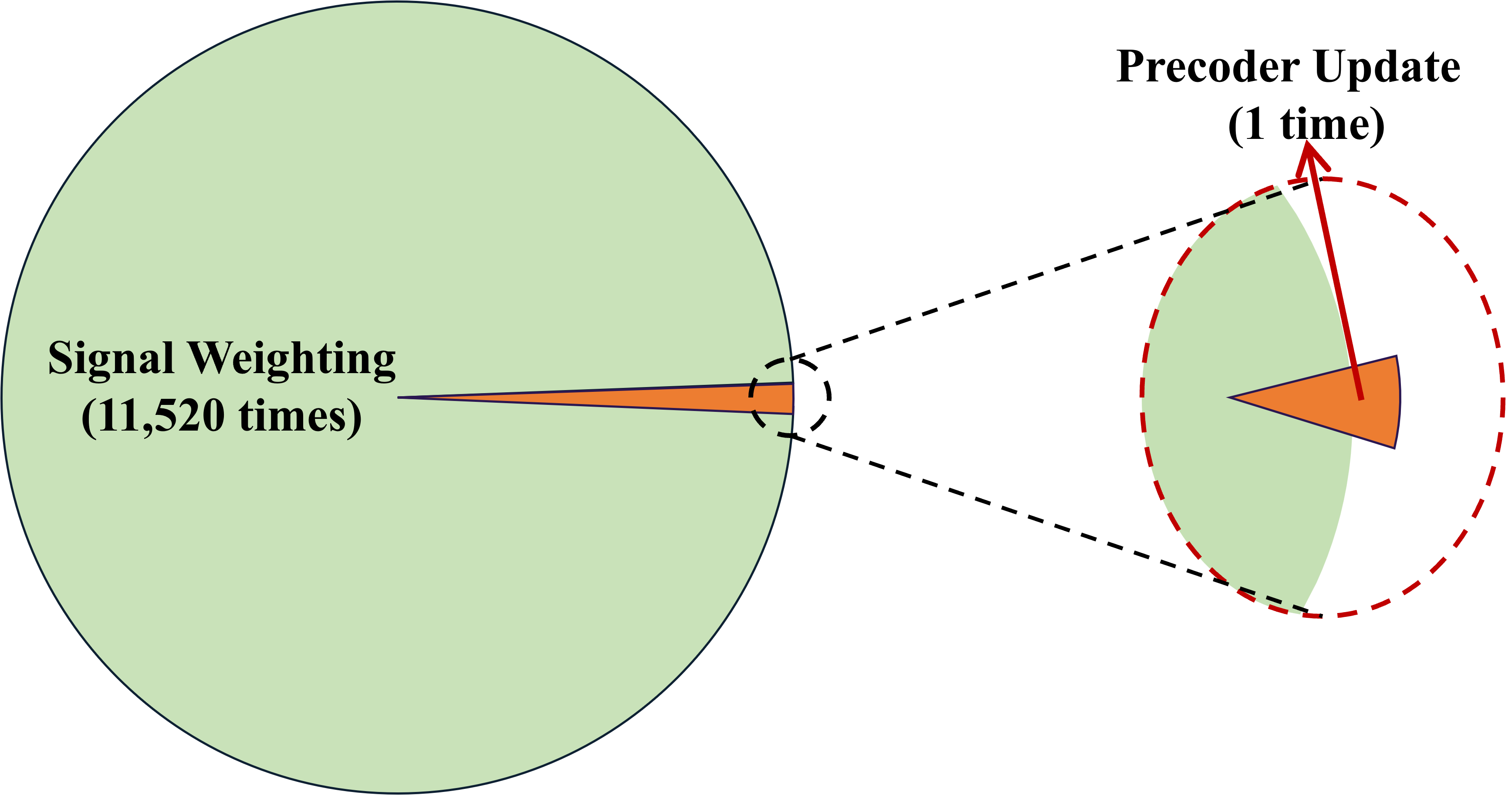}
\caption{Execution frequency ratio of signal weighting to precoder update within a single channel coherence interval.}
\label{piechart}
\end{figure}

\begin{itemize}
\item \textbf{Signal Weighting Cost-Aware Precoding Formulation and Dual WSR-Driven Row-Sparsity Models:}
We formulate a WSR-driven sparse digital precoding problem for fully-digital MIMO systems from the signal weighting cost perspective. To reduce the baseband computational overhead, we propose two spatial sparsity models for the precoding matrix: 1) an {angle-level sparsity} model, which enforces a {joint row-sparsity} structure, and 2) a more flexible {angle-user-level sparsity} model, characterized by a {user-specific row-sparsity} pattern that allows each user to select an individual set of beams, thereby substantially increasing the design freedom of sparse precoding.
\end{itemize}

\begin{itemize}
    %\item \textbf{Optimal Low-Dimensional Subspace Structures and Cubic-to-Linear Complexity Reduction:} The core theoretical contribution of this work is the rigorous mathematical proof of the optimal precoder structures under the proposed sparsity models. We reveal that, despite the highly non-convex mixed-integer non-linear programming (MINLP) nature of the formulated problems, the optimal sparse precoding matrices for both models intrinsically lie within specific low-dimensional subspaces defined by the effective channel matrices. This fundamental theoretical insight enables us to strictly reduce the dimensionality of the continuous optimization variables from the massive number of BS antennas $M$ to the much smaller total number of data streams $N$ ($N \ll M$). Consequently, this breakthrough dramatically reduces the computational complexity of the precoder design from cubic with respect to the number of antennas, $\mathcal{O}(M^3)$, to merely linear, $\mathcal{O}(MN^2)$. Furthermore, the discovered subspace structures can be readily extended to other sparse precoding scenarios, such as multi-cell cooperation, to effectively reduce problem dimensionality and algorithm complexity.

    \item \textbf{\textcolor{black}{Optimal Low-Dimensional Subspace Characterization for Sparse Precoding:}}
We rigorously show that, despite the  non-convex mixed-integer non-linear programming (MINLP) nature of the formulated problems, the optimal sparse precoder under both proposed sparsity models lies in a low-dimensional subspace determined by the effective channel matrices. \textcolor{black}{This structural result reduces the dimensionality of the continuous optimization variables from the number of BS antennas $M$ to the aggregate receive dimension $N=\sum_{k=1}^{K}N_k$, which is much smaller than $M$ in the considered massive MIMO setting.} \textcolor{black}{Compared with conventional sparsity-constrained WMMSE precoding operating in the full-dimensional precoder space, the resulting reformulation reduces the per-iteration precoder-design complexity from $\mathcal{O}(M^3)$ to $\mathcal{O}(MN^2)$.} %Furthermore, the discovered subspace structures can be readily extended to other sparse precoding scenarios, such as multi-cell cooperation, to effectively reduce problem dimensionality and algorithm complexity.
\end{itemize}

\begin{itemize}
    %\item \textbf{WSR-Based Formulation and Optimal Subspace Structure for Angle-Level Sparsity:}
    %We formulate, for the time, the sparse digital precoding problem for all-digital MIMO systems from a WSR maximization perspective, explicitly targeting the signal weighting cost bottleneck. We first consider an angle-level sparsity model, in which all users share  a common set of angle-domain basis vectors (beams). A key  contribution is the mathematical proof that the optimal precoder under this model lies in a specific low-dimensional subspace defined by the channel matrices. This result enables a reformulation of  the original mixed-integer non-linear program (MINLP) problem, reducing the dimensionality of the optimization variables from the number of antennas $M$ to the total number of data streams $N$, where $N \ll M$.

    %\item \textbf{Novel Angle-User-Level Sparsity Model and Subspace Characterization:}
    %We further introduce a more flexible and practically relevant angle-user-level sparsity model, in which each user selects an individual  set of beams. This model substantially increases the degrees of freedom for sparse precoding  design. We  rigorously show that this more general formulation  also admits an optimal low-dimensional subspace structure, thereby demonstrating the robustness and generality of the proposed  theoretical framework.

    \item \textbf{Baseband-Efficient  WMMSE Algorithms:}
Leveraging the low-dimensional subspace structures under the two sparsity models, we develop two baseband-efficient WMMSE algorithms, namely the angle-level low-complexity low-dimensional sparse precoding (ALLSP) algorithm and the angle-user-level low-complexity low-dimensional sparse precoding (AULLSP) algorithm. For both algorithms, the combinatorial beam selection subproblem is handled via a concave penalty-based relaxation and a majorize-minimization (MM) procedure, which transforms the original intractable problem into a simple linear program admitting a closed-form solution through a sorting operation.

    \item \textbf{Performance Validation and Complexity Analysis:}
    \textcolor{black}{Through complexity analysis and simulations, we demonstrate that the proposed framework achieves WSR performance close to that of the full-dimensional WMMSE benchmark, while substantially reducing the signal-weighting cost relative to dense precoding.} \textcolor{black}{Moreover, compared with conventional sparsity-constrained WMMSE precoding, i.e., S-WMMSE, the proposed low-dimensional reformulation substantially reduces the computational complexity of precoder design.}
\end{itemize}
The remainder of this paper is organized as follows. Section II introduces the system model. Sections III and  IV detail the design and optimization algorithms for angle-level and angle-user-level sparse precoding, respectively. Section V provides an  analysis of the computational and signal weighting complexities. Section VI validates the proposed methods through simulations. Finally, Section VII concludes the paper.

\section{Downlink Massive MIMO Transmission Model}

\subsection{Downlink Transmission in the Antenna Domain}
Consider a downlink MU-MIMO communication system consisting of a BS equipped with $M$ transmit antennas and $K$ geographically distributed users. Each user terminal $k \in \{1, \ldots, K\}$ is equipped with $N_k$ receive antennas. Let $\mathbf{s}_k \in \mathbb{C}^{D_k \times 1}$ denote  the data symbol vector intended for user $k$, where $D_k$ represents the number of spatially multiplexed data streams allocated to this user. The total number of data streams transmitted  across all users is given by $D \triangleq \sum_{k=1}^K D_k$. The   data symbols are modeled as independent circularly symmetric complex Gaussian vectors, i.e., $\mathbf{s}_k \sim \mathcal{CN}(\mathbf{0}, \mathbf{I})$ for all $k \in \{1, \ldots, K\}$. 

The channel matrix between the BS and user $k$ in the antenna domain is denoted by $\mathbf{H}_{\text{ant},k} \in \mathbb{C}^{N_k \times M}$. Accordingly, the  received signal at user $k$ can be expressed as:
\begin{align}
\mathbf{y}_{k} = \underbrace{\mathbf{H}_{\text{ant},k}\mathbf{P}_{\text{ant},k}\mathbf{s}_k}_{\text{desired signal}} + \underbrace{\sum_{j=1,j\neq k}^K\mathbf{H}_{\text{ant},k}\mathbf{P}_{\text{ant},j}\mathbf{s}_j}_{\text{interference}} + \mathbf{n}_k, \label{eqn:received_signal_model}
\end{align}
where $\mathbf{P}_{\text{ant},k} \in \mathbb{C}^{M \times D_k}$ denotes the precoding matrix for user $k$ in the antenna domain, and $\mathbf{n}_k \in \mathbb{C}^{N_k \times 1}$ 
represents  the additive white Gaussian noise (AWGN) vector at user $k$, which is modeled as $\mathbf{n}_k \sim \mathcal{CN}(\mathbf{0}, \sigma_k^2\mathbf{I}_{N_k})$. Based on \eqref{eqn:received_signal_model}, the achievable rate of user $k$ is given by:
\begin{align}
\label{r_define}
R_k = &\log\det \bigg(\mathbf{I}_{N_k} + \mathbf{H}_{\text{ant},k}\mathbf{P}_{\text{ant},k}\mathbf{P}_{\text{ant},k}^H\mathbf{H}_{\text{ant},k}^H \nonumber \\
&\Big(\sum_{j\neq k}\mathbf{H}_{\text{ant},k}\mathbf{P}_{\text{ant},j}\mathbf{P}_{\text{ant},j}^H\mathbf{H}_{\text{ant},k}^H+\sigma_k^2\mathbf{I}_{N_k}\Big)^{-1}\bigg).
\end{align}
A common objective is to maximize the system spectral efficiency by solving the following WSR maximization problem:
\begin{subequations}
\label{int_pro}
\begin{align}
\max_{\mathbf{P}_{\text{ant}}} \quad &\sum_{k=1}^K\alpha_k R_k \\
\text{s.t.} \quad &\sum_{k=1}^K\text{Tr}\left(\mathbf{P}_{\text{ant},k}\mathbf{P}_{\text{ant},k}^H\right)\leq P_{\max},
\end{align}
\end{subequations}
where $\mathbf{P}_{\text{ant}}\triangleq [\mathbf{P}_{\text{ant},1}, \ldots, \mathbf{P}_{\text{ant},K}]$ denotes the aggregate precoding matrix, $\alpha_k$ represents the weight assigned to user $k$, and $P_{\max}$ is the total transmit power constraint. Solving problem \eqref{int_pro} typically requires iterative algorithms \cite{christensen2008weighted,shi2011iteratively}. Consequently, most prior studies focus on designing low-complexity algorithms that yield suboptimal yet effective precoders. However, emphasizing only the complexity of computing the precoder neglects a crucial implementation aspect, i.e., the computational cost of applying the precoder to the data symbols. In block precoding schemes, the corresponding matrix-vector multiplication (hereafter referred to as signal weighting) must be performed for every symbol vector in the block. The high execution frequency of this operation introduces a substantial computational bottleneck in practical systems.

\subsection{Precoding Computation Complexity Versus Signal Weighting Cost}
To justify the necessity of sparse precoding in a fully-digital architecture, it is essential to distinguish between the one-time complexity of precoder derivation and the recurring overhead of its real-time application.

The transmitted signal at the BS is given by the product of the precoding matrix and the data symbols, i.e., $\mathbf{x} = \mathbf{P}_{\text{ant}}\mathbf{s}$, where $\mathbf{s}\triangleq [\mathbf{s}_1^T, \ldots, \mathbf{s}_K^T]^T \in\mathbb{C}^{D\times 1}$. 
Note that the timescale for computing $\mathbf{P}_{\text{ant}}$ and $\mathbf{x}$ is significantly different in practice. The precoding matrix $\mathbf{P}_{\text{ant}}$ is typically computed once per channel coherence interval and remains fixed over multiple time-frequency resource blocks (RBs), whereas the data symbol vector $\mathbf{s}$ changes at every transmission instance. As a result, although applying the precoder involves  a simple matrix-vector multiplication, this operation must be executed repeatedly at a very high rate. 
The cumulative computational cost of signal weighting may exceed the computational complexity of computing the precoding matrix $\mathbf{P}_{\text{ant}}$ a single time.

\begin{figure}[t] 
\centering
\hspace*{-0.04\textwidth} % 可调节偏移量
\includegraphics[width=0.45\textwidth]{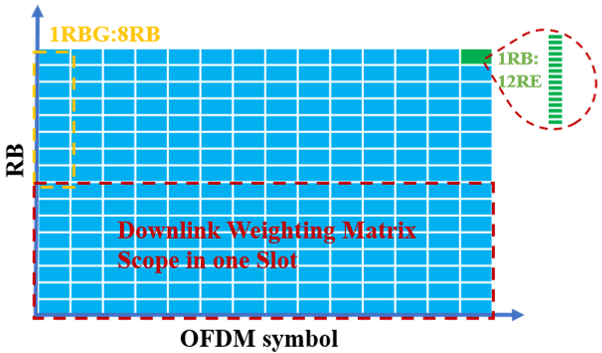}
\caption{Downlink 8RB weighting matrix scope in one slot}
\label{weightscope}
\vspace{-1.2em}
\end{figure}

To illustrate this effect, consider  a representative 5G system, in which the precoder is computed once per channel sounding period (e.g., 5 ms) and then applied to a user scheduled over 8 RBs, as illustrated in Fig. \ref{weightscope}. Assuming a slot duration of 0.5 ms, this period contains 10 slots.  Each RB consists of 12 subcarriers, and each slot includes  approximately 12 data-bearing Orthogonal Frequency Division Multiplexing (OFDM) symbols after accounting for control and reference signals. The total number of resource elements requiring precoding is therefore: 
$\text{8} \, (\text{RBs}) \times \text{12} \, (\text{subcarriers}) \times \text{10} \, (\text{slots}) \times \text{12} \, (\text{OFDM symbols}) = \text{11,520}.$
This implies that the same precoding matrix must be applied through 11,520 distinct matrix–vector multiplications within a single channel coherence period, resulting in a substantial computational burden in the baseband processor.
Furthermore, 5 ms corresponds to an ideal SRS periodicity. In congested networks, this period can be much longer due to pilot shortages. For example, increasing the SRS periodicity to \textcolor{black}{80 ms} raises the number of signal weighting operations per precoder update to \textcolor{black}{184,320}, further aggravating the baseband computational burden.

\subsection{Downlink Massive MIMO Transmission in the Angle Domain}
\begin{figure}[!t]
\centering
\hspace{-1.2em}
\subfigure[]{
\includegraphics[width=0.19\textwidth]{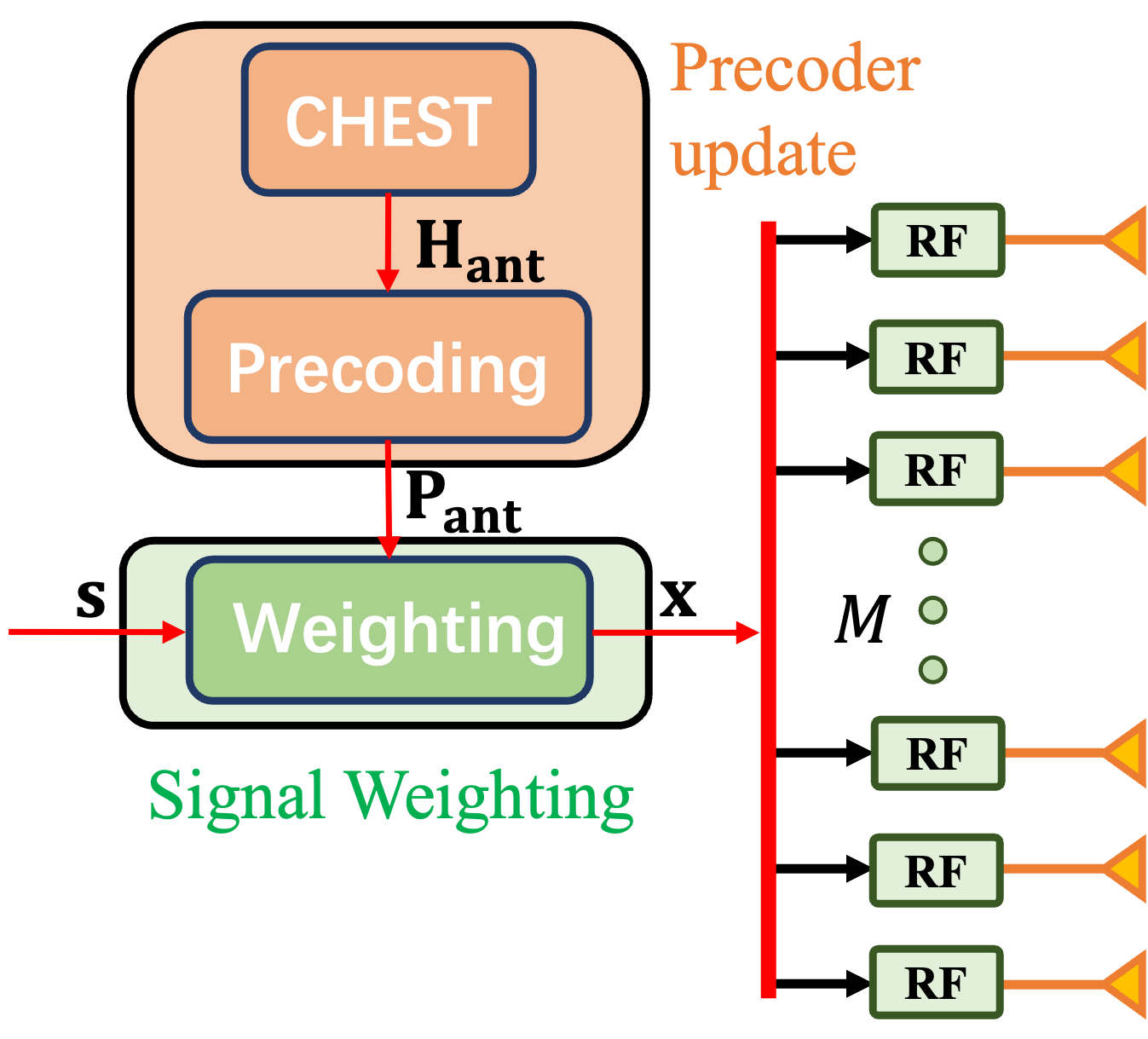}
%\label{}
}
%\hspace{0.7em}
\subfigure[]{
\includegraphics[width=0.25\textwidth]{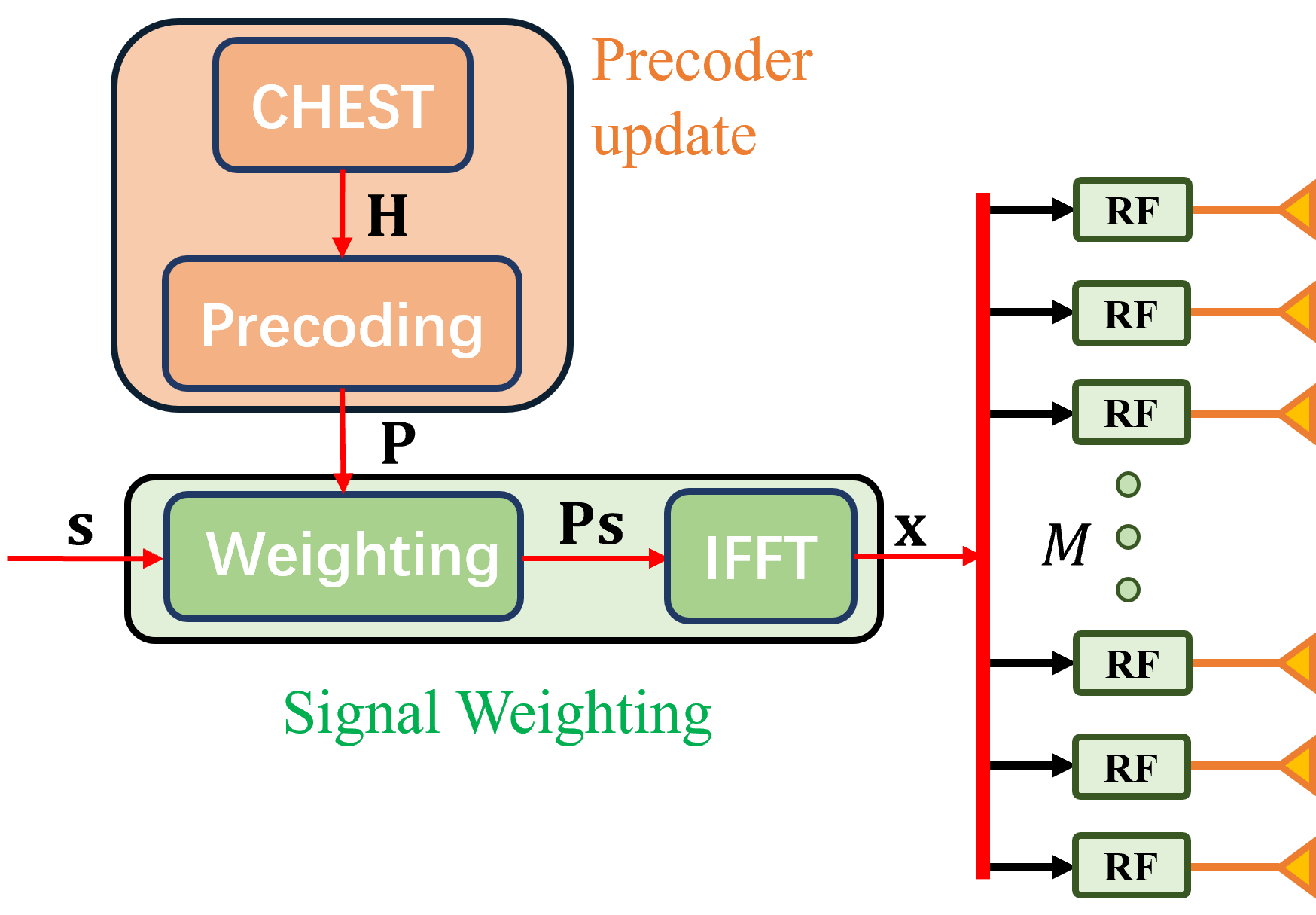}
%\label{}
}
\caption{Downlink baseband processing architectures: (a)  antenna domain (b) angle domain.}
\label{basebandprocess1}
\end{figure}

As discussed above, it is essential to account not only for the complexity of computing the precoding matrix but also for the prohibitive computational cost incurred during the signal weighting stage. A natural approach to reducing this high cost is to directly design a sparse antenna-domain precoding matrix, $\mathbf{P}_{\text{ant}}$. If $\mathbf{P}_{\text{ant}}$ contains a large number of zero entries, the number of multiplications required to compute the transmitted signal $\mathbf{x} = \mathbf{P}_{\text{ant}}\mathbf{s}$ can be substantially reduced.

However, this direct approach encounters a fundamental limitation: the physical wireless channel in the antenna domain is typically dense. Imposing sparsity on $\mathbf{P}_{\text{ant}}$ without an appropriate basis transformation  leads to a  mismatch with the  channel structure, which can result in severe performance degradation. Therefore,  it is more effective to operate in a domain in which the channel is inherently sparse.

For massive MIMO systems with large antenna arrays, the angle domain provides this desirable property\cite{gonultacs2021hardware,tong2023low}. Due to limited scattering, the channel energy is concentrated  in a small number of dominant angular directions, making the angle domain a natural and efficient basis for sparse precoder design.

As illustrated in Fig. \ref{basebandprocess1}, downlink baseband processing can be represented either in the antenna domain (Fig. \ref{basebandprocess1}(a)) or the angle domain (Fig. \ref{basebandprocess1}(b)). The angle-domain channel \( \mathbf{H}_k \) for user $k$ is obtained by applying a discrete  Fourier transform (DFT) to the antenna-domain channel \( \mathbf{H}_{\text{ant},k} \), i.e.,
\[
\mathbf{H}_k = \mathbf{H}_{\text{ant},k} \mathbf{F}_M^H,
\]
where \( \mathbf{F}_M \in \mathbb{C}^{M \times M} \) denotes the unitary DFT matrix\cite{gonultacs2021hardware,tong2023low}. Accordingly, the antenna-domain precoder $\mathbf{P}_{\text{ant},k}$ can be expressed in terms of its angle-domain representation $\mathbf{P}_k$ as $\mathbf{P}_{\text{ant},k} = \mathbf{F}_M^H \mathbf{P}_k$, as shown in Fig. \ref{basebandprocess1}(b). Substituting these expressions into the original signal model in \eqref{eqn:received_signal_model} yields an equivalent angle-domain representation of the downlink transmission:
\begin{align}
\mathbf{y}_{k} &= (\mathbf{H}_k \mathbf{F}_M) (\mathbf{F}_M^H \mathbf{P}_k) \mathbf{s}_k + \sum_{j \neq k} (\mathbf{H}_k \mathbf{F}_M) (\mathbf{F}_M^H \mathbf{P}_j) \mathbf{s}_j + \mathbf{n}_k \nonumber \\
&= \mathbf{H}_k \mathbf{P}_k \mathbf{s}_k + \sum_{j \neq k} \mathbf{H}_k \mathbf{P}_j \mathbf{s}_j + \mathbf{n}_k.
\end{align}
Since the angle-domain channel $\mathbf{H}_k$ \textcolor{black}{is approximately sparse}\cite{gonultacs2021hardware,tong2023low}, a sparse precoding matrix $\mathbf{P}_k$ can be designed to primarily operate on the dominant angular components of the channel while setting the remaining elements to zero. This sparsity in $\mathbf{P}_k$ substantially reduces the complexity of the signal weighting operation. Specifically, the transmitted signal can be efficiently computed as $\mathbf{x} = \mathbf{P}_{\text{ant}}\mathbf{s} = \mathbf{F}_M^H (\mathbf{P}\mathbf{s})$, where the inner multiplication, $\mathbf{P}\mathbf{s}$, has  very low complexity due to the sparsity of $\mathbf{P}$, and the multiplication by $\mathbf{F}_M^H$ can be efficiently  implemented using the inverse fast Fourier transform (IFFT).

To effectively translate the inherent channel sparsity into a reduction in signal weighting cost, the structure of the non-zero entries in the precoding matrix $\mathbf{P}$ must be carefully defined. Since the rows of the angle-domain precoder correspond to the transmit angular beams, imposing sparsity is equivalent to selecting a subset of active beams. This selection can be enforced either globally to minimize the aggregate dimension of the optimization variables, or individually for each user to maximize the degrees of freedom matching their specific channel directions. Motivated by this trade-off between implementation simplicity and system performance, we consider two sparse precoding strategies:

\begin{itemize}
\item an \textbf{angle-level sparse precoding} scheme, where a common set of beams is selected for all users to induce a row-sparse structure on $\mathbf{P}$; 
\item an \textbf{angle-user-level sparse precoding} scheme, where the set of active beams is optimized specifically for each user to exploit multi-user diversity.
\end{itemize}

\begin{figure}[!t]
\centering
\hspace{-1.2em}
\subfigure[]{
\includegraphics[width=0.2\textwidth]{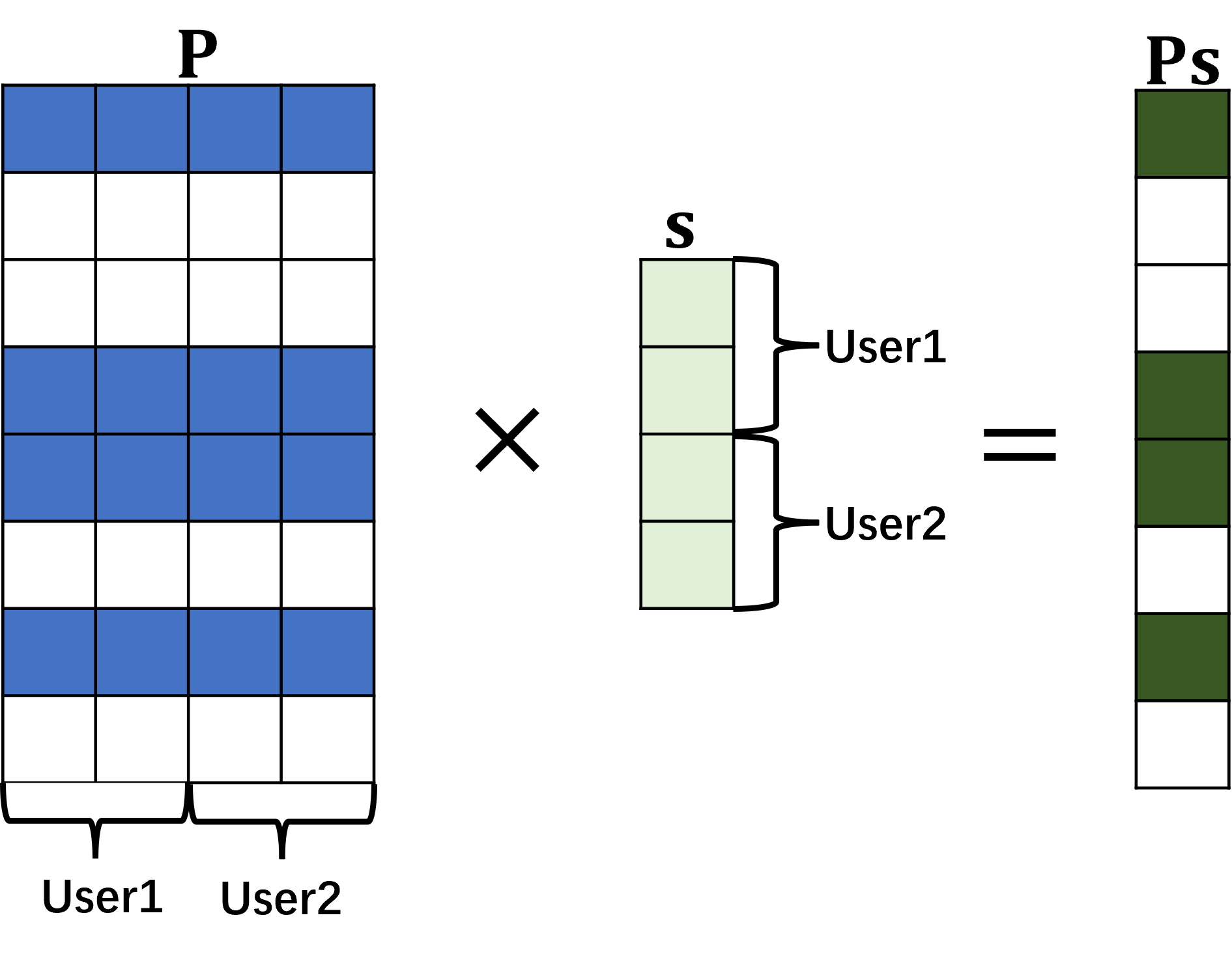}
%\label{}
}
\hspace{1.2em}
\subfigure[]{
\includegraphics[width=0.2\textwidth]{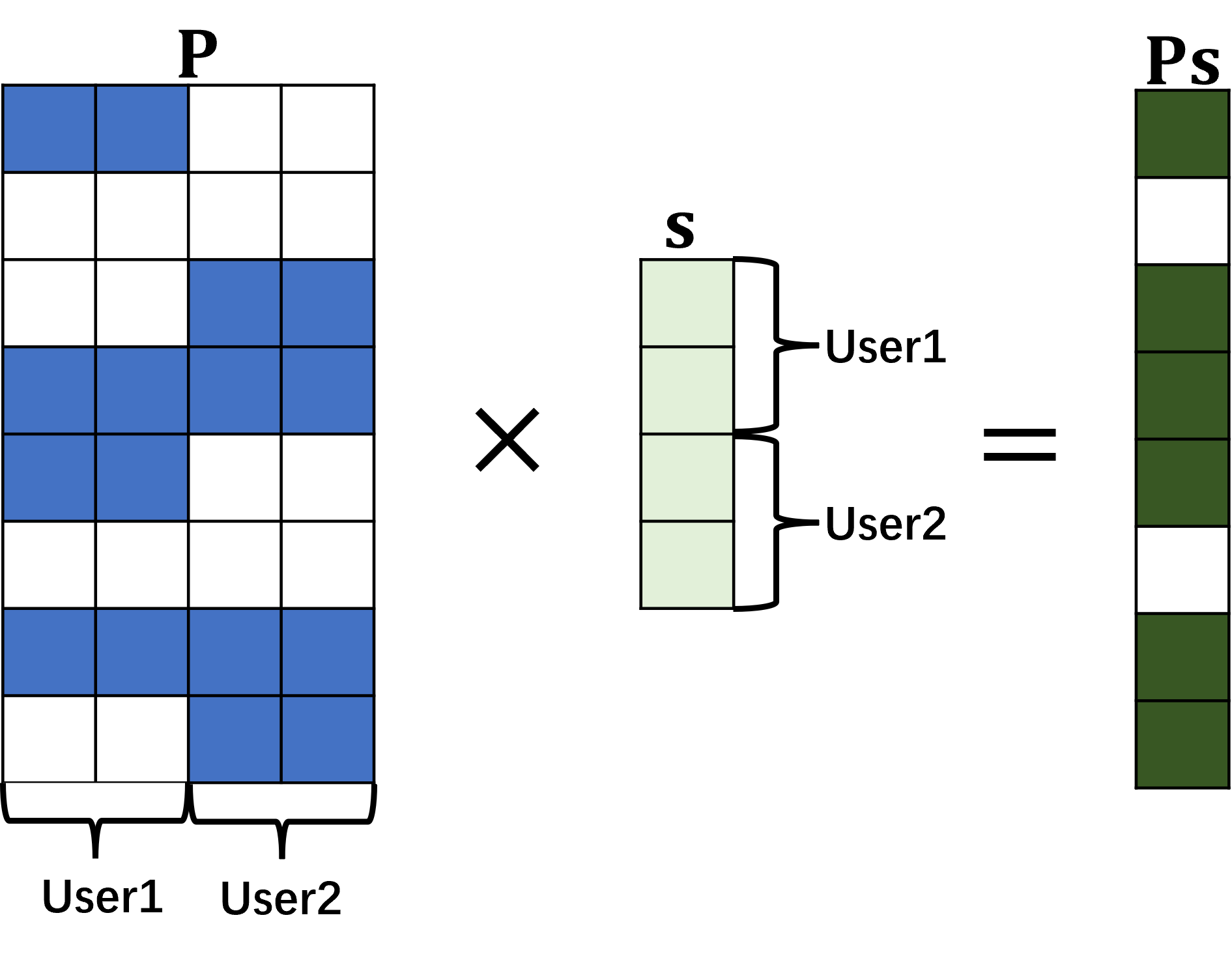}
%\label{}
}
\caption{Downlink sparse precoding matrices: (a) angle-level sparsity and (b) angle-user-level sparsity. Dark-shaded squares in the  $\mathbf{P}$ and $\mathbf{P}\mathbf{s}$ sequences indicate non-zero entries, whereas light-shaded ones correspond to zero entries.}
%\hspace{-1.2em}
\label{sparsematrix}
\vspace{-1.2em}
\end{figure}

These strategies are illustrated by the sparse precoding matrices in Fig. \ref{sparsematrix}(a) and Fig. \ref{sparsematrix}(b), respectively.

\begin{remark}
\textbf{(Comparison with Existing Schemes)}
Existing schemes for reducing signal weighting cost \cite{gonultacs2021hardware,tong2023low} \textcolor{black}{do not directly optimize the sum rate}. By contrast, the proposed framework \textcolor{black}{jointly optimizes beam selection and precoding coefficients under the WSR criterion for both angle-level and angle-user-level sparsity}, and \textcolor{black}{exploits the corresponding optimal low-dimensional subspace structures} for low-complexity optimization.
\end{remark}

\section{Downlink Angle-Level Sparse Precoding Problem}
\subsection{Problem Formulation} 
We first address the downlink angle-level sparse precoding problem in this section, while the downlink angle-user-level sparse precoding problem is discussed in Section IV. In this section, we first mathematically formulate the target problem, and then reformulate it into a low-dimensional one based on the subspace characterization of optimal angle-domain precoding matrices. Finally, we propose a low-complexity low-dimensional sparse precoding algorithm under the WMMSE framework.

%As discussed in the previous section, the sparsity structure of the massive MIMO channel is inherently captured in the angle domain. To mitigate  the substantial computational overhead associated with signal weighting, we propose an angle-level sparse precoding strategy. Unlike conventional antenna selection schemes that deactivate physical antenna elements, the proposed approach selects a subset of dominant angular beams for  transmission with the rows of the angle-domain precoding matrix associated with inactive angular bins set to zero.

We consider an angle-level sparse precoding design, where only a subset of angular beams is activated for transmission. Let $\mathbf{P} \in \mathbb{C}^{M \times D}$ denote the angle-domain precoding matrix. The proposed design enforces a row-sparse structure on $\mathbf{P}$, where the number of nonzero rows is strictly limited. This constraint can be expressed as:
\begin{equation}
    \sum_{i=1}^{M} \operatorname{sgn}(\|\mathbf{P}_{i,:}\|_2) = K_s,
\end{equation}
where $\mathbf{P}_{i,:}$ denotes the $i$-th row of $\mathbf{P}$, $\|\cdot\|_2$ is the $\ell_2$-norm, and $K_s$ indicates the target number of active angular  beams. This specific structure corresponds to group sparsity, where each group is defined by a row of the precoding matrix that jointly serves  all users within that group.

Our objective is to maximize the system WSR subject to the total transmit power constraint and the aforementioned row-sparsity constraint. The sparse precoding problem is formulated as:
\begin{subequations} \label{prob:original_sparse}
\begin{align}
    \max_{\mathbf{P}} \quad & \sum_{k=1}^{K} \alpha_{k} R_k(\mathbf{P}) \\
    \text{s.t.} \quad & \|\mathbf{P}\|_F^2 \le P_{max}, \\
    & \sum_{i=1}^{M} \operatorname{sgn}(\|\mathbf{P}_{i,:}\|_2) = K_s, \label{eq:constraint_l0}
\end{align}
\end{subequations}
where $R_k(\mathbf{P})$ represents the achievable rate of user $k$ as defined in \eqref{r_define}.

Problem \eqref{prob:original_sparse} is a non-convex MINLP problem due to the combinatorial nature of the row-support constraint \eqref{eq:constraint_l0} and the non-convex objective function. Although group-sparse optimization has been extensively studied, often through convex relaxations such as mixed $\ell_2/\ell_1$-norm regularization or iterative reweighted algorithms\cite{hong2013joint}, directly applying these techniques to massive MIMO systems poses substantial challenges. In particular, standard convex solvers or reweighted minimization algorithms typically exhibit computational complexity that scales cubically with the number of transmit antennas, i.e., $\mathcal{O}(M^3)$.  Therefore, it is crucial to develop a design that achieves substantial weighting cost reduction without incurring excessive computational complexity during precoding computation.

%Although group-sparse optimization has been extensively studied, often through  mixed $\ell_2/\ell_1$-norm regularization, directly applying these techniques to massive MIMO systems poses substantial challenges. In particular, standard  convex relaxations or iterative reweighted algorithms typically exhibit computational complexity that scales cubically with the number of transmit antennas ,i.e., $\mathcal{O}(M^3)$. Since the primary motivation for introducing sparsity is to reduce the computational burden of the signal weighting stage, employing a high-complexity algorithm to generate the precoder is counterproductive. Therefore, it is crucial to develop a design that achieves substantial weighting cost reduction without incurring excessive computational complexity during precoding computation.

To obtain an efficient formulation, we introduce a diagonal binary selection matrix $\boldsymbol{\Delta} \in \textcolor{black}{\mathbb{R}}^{M \times M}$, where the diagonal entry $[\boldsymbol{\Delta}]_{ii} \in \{0, 1\}$ indicates whether the $i$-th angular beam is active. The angle-domain precoder is then decomposed as $\mathbf{P} = \boldsymbol{\Delta}\mathbf{V}$, where $\mathbf{V} \triangleq [\mathbf{V}_1, \ldots, \mathbf{V}_K] \in \mathbb{C}^{M \times D}$ contains the corresponding precoding coefficients. Substituting this decomposition into the angle-domain signal model, the received signal at user $k$ is given by:
\begin{align}
\mathbf{y}_k = \mathbf{H}_k\boldsymbol{\Delta}\mathbf{V}_k\mathbf{s}_k + \sum_{j \neq k} \mathbf{H}_k\boldsymbol{\Delta}\mathbf{V}_j\mathbf{s}_j + \mathbf{n}_k,
\label{eq:sparse_signal_model}
\end{align}
where $\mathbf{H}_k$ is the angle-domain channel defined previously. Since the DFT matrix $\mathbf{F}_M$ is unitary, the total transmit power constraint is preserved across domains, i.e., $\|\mathbf{P}_{\text{ant}}\|_F^2 = \|\mathbf{F}_M^H \boldsymbol{\Delta}\mathbf{V}\|_F^2 = \|\boldsymbol{\Delta}\mathbf{V}\|_F^2$.

Accordingly, the WSR maximization problem  subject to angle-level sparsity constraints is formulated as:
\begin{subequations}
\label{pro:anglesparse}
\begin{align}
\max_{\{\mathbf{V}_k\}, \boldsymbol{\Delta}} \quad & \sum_{k=1}^K \alpha_k \log \det \Big( \mathbf{I}_{N_k} + \mathbf{H}_k\boldsymbol{\Delta}\mathbf{V}_k\mathbf{V}_k^H\boldsymbol{\Delta}^H\mathbf{H}_k^H \nonumber\\
& \times \big( \sum_{j \neq k} \mathbf{H}_k\boldsymbol{\Delta}\mathbf{V}_j\mathbf{V}_j^H\boldsymbol{\Delta}^H\mathbf{H}_k^H + \sigma_k^2\mathbf{I}_{N_k} \big)^{-1} \Big) \\
\text{s.t.} \quad & \sum_{k=1}^K \mathrm{Tr}(\boldsymbol{\Delta}\mathbf{V}_k\mathbf{V}_k^H\boldsymbol{\Delta}^H) \leq P_{\max}, \\
& [\boldsymbol{\Delta}]_{ii} \in \{0,1\}, \quad \forall i \in \{1, \ldots, M\}, \\
& \sum_{i=1}^M [\boldsymbol{\Delta}]_{ii} = K_s.
\end{align}
\end{subequations}
Problem \eqref{pro:anglesparse} involves  continuous variables $\{\mathbf{V}_k\}$ and discrete binary variables $\boldsymbol{\Delta}$, and is therefore classified  as a MINLP problem. Such problems are inherently NP-hard and generally intractable to solve directly. To address this challenge, we first analyze the structure of the optimal solution in order to identify the underlying low-dimensional subspace in which the optimal precoder resides. Based on this structural insight, we then develop an efficient low-complexity algorithm for solving the problem.

\subsection{Optimal Sparse Precoder Structure and Problem Reformulation}
\begin{proposition}[Subspace Characterization]\label{prop:subspace}
The optimal angle-domain sparse precoding matrix $\boldsymbol{\Delta}^*\mathbf{V}_k^*$ lies in the subspace spanned by $\boldsymbol{\Delta}^*\mathbf{H}^H$. Specifically, it satisfies the following structural relationship:
\begin{align}
\boldsymbol{\Delta}^*\mathbf{V}_k^* = \boldsymbol{\Delta}^*\mathbf{H}^H\mathbf{X}_k^*,
\label{eq:precoder_subspace}
\end{align}
where $\mathbf{H} \triangleq [\mathbf{H}_1^T, \ldots, \mathbf{H}_K^T]^T \in \mathbb{C}^{N \times M}$ denotes the aggregate angle-domain channel matrix with $N = \sum_{k=1}^K N_k$, and $\mathbf{X}_k \in \mathbb{C}^{N \times D_k}$ represents the low-dimensional projection coefficient matrix. The proof is provided in Appendix A.
\end{proposition}
% \begin{proof}
%    See Appendix A.
% \end{proof}
Proposition \ref{prop:subspace} indicates that the design of the high-dimensional sparse precoder $\boldsymbol{\Delta}\mathbf{V}_k$ can be equivalently reduced to determining the low-dimensional projection weights $\{\mathbf{X}_k\}$ together with the binary selection matrix $\boldsymbol{\Delta}$. Leveraging the structural result, we substitute the parametric representation $\boldsymbol{\Delta}\mathbf{V}_k = \boldsymbol{\Delta}\mathbf{H}^H\mathbf{X}_k$ into \eqref{pro:anglesparse}, thereby transforming the original problem into the following equivalent optimization formulation:

\begin{subequations}
\begin{align}
\max_{\{\mathbf{X}_k\}, \boldsymbol{\Delta}} \quad & \sum_{k=1}^K \alpha_k\log \det ( \mathbf{I} + \mathbf{H}_k\boldsymbol{\Delta}\mathbf{H}^H\mathbf{X}_k\mathbf{X}_k^H\mathbf{H}\boldsymbol{\Delta}\mathbf{H}_k^H  \nonumber\\
&   ( \sum_{j \neq k} \mathbf{H}_k\boldsymbol{\Delta}\mathbf{H}^H\mathbf{X}_j\mathbf{X}_j^H\mathbf{H}\boldsymbol{\Delta}\mathbf{H}_k^H + \sigma_k^2\mathbf{I} )^{-1} ) \\
\mathrm{s.t.} \quad & \sum_{k=1}^K \mathrm{Tr}(\mathbf{H}\boldsymbol{\Delta}\mathbf{H}^H\mathbf{X}_k\mathbf{X}_k^H) \leq P_{max}, \label{eq:power_constraint_1} \\
& [\boldsymbol{\Delta}]_{ii} \in \{0,1\},\ \forall i, \\
& \sum_{i=1}^M [\boldsymbol{\Delta}]_{ii} = K_s.
\end{align}
\label{eq:equivalent_problem1}
\end{subequations}

To handle the sum-power constraint \eqref{eq:power_constraint_1} in \eqref{eq:equivalent_problem1}, we adopt a scale-invariant reformulation, in which the power budget is absorbed into the noise regularization term. The resulting equivalent problem is given by:
\begin{subequations}
\begin{align}
\max_{\{\mathbf{X}_k\}, \boldsymbol{\Delta}} \quad & \sum_{k=1}^K \alpha_k\log \det ( \mathbf{I} + \mathbf{H}_k\boldsymbol{\Delta}\mathbf{H}^H\mathbf{X}_k\mathbf{X}_k^H\mathbf{H}\boldsymbol{\Delta}\mathbf{H}_k^H  \nonumber\\
&   ( \sum_{j \neq k} \mathbf{H}_k\boldsymbol{\Delta}\mathbf{H}^H\mathbf{X}_j\mathbf{X}_j^H\mathbf{H}\boldsymbol{\Delta}\mathbf{H}_k^H + \mathbf{N}_k )^{-1} ) \\
\mathrm{s.t.} \quad 
& [\boldsymbol{\Delta}]_{ii} \in \{0,1\},\ \forall i, \\
& \sum_{i=1}^M [\boldsymbol{\Delta}]_{ii} = K_s.
\end{align}
\label{pro:equivalent_problem2}
\end{subequations}

\noindent where the effective noise covariance matrix is defined as $\mathbf{N}_k = \frac{\sigma_k^2}{P_{\max}} \big( \sum_{i=1}^{K} \mathrm{Tr}(\mathbf{X}_i^H \mathbf{H}\boldsymbol{\Delta}\mathbf{H}^H \mathbf{X}_i) \big) \mathbf{I}_{N_k}$. The equivalence between \eqref{eq:equivalent_problem1} and \eqref{pro:equivalent_problem2} follows from a simple scaling argument\cite{hu2020iterative}: the optimal solution of \eqref{pro:equivalent_problem2} can be properly rescaled to satisfy the sum-power constraint in \eqref{eq:equivalent_problem1} with equality, while such a common rescaling does not affect the objective value of \eqref{pro:equivalent_problem2}. Due to space limitations, the detailed proof is omitted here. Hence, the optimal sparse precoding matrix for the original problem can be recovered as
\begin{equation}
    \boldsymbol{\Delta}^*\mathbf{V}_k^* = \sqrt{\omega} \boldsymbol{\Delta}^*\mathbf{H}^H \mathbf{X}_k^*,
\end{equation}
where $\omega$ denotes the scaling factor selected such that the sum-power constraint is satisfied with equality.

Although the problem dimensionality has been significantly reduced,  problem \eqref{pro:equivalent_problem2} 
remains a challenging MINLP problem due to the binary constraints on $\boldsymbol{\Delta}$. To address this challenge, we propose the ALLSP algorithm in the next subsection to obtain a high-quality solution efficiently.

\subsection{Angle-Level Low-Complexity Low-Dimensional Sparse Precoding  Algorithm Design}
Exploiting the equivalence between the WSR and the WMMSE problem \cite{shi2011iteratively}, we reformulate the optimization problem into an equivalent WMMSE form. By introducing auxiliary variables, the receiver filters $\mathbf{U}=\{\mathbf{U}_k\}_{k=1}^K$ and weight matrices $\mathbf{W}=\{\mathbf{W}_k\}_{k=1}^K$,we obtain the following equivalent optimization problem:
\begin{subequations} \label{eq:wmmse_reform_corrected}
\begin{align}
    \min_{\{\mathbf{W}_k\},\{\mathbf{U}_k\}, \{\mathbf{X}_k\},\boldsymbol{\Delta}}  & \sum_{k=1}^K \alpha_k\left[ \mathrm{Tr}(\mathbf{W}_k\mathbf{E}_k) - \log\det(\mathbf{W}_k) \right] \label{eq:wmmse_reform_corrected_a}\\
    \mathrm{s.t.} \quad & [\boldsymbol{\Delta}]_{ii} \in \{0,1\}, \quad \forall i, \\
    & \sum_{i=1}^M [\boldsymbol{\Delta}]_{ii} = K_s.
\end{align}
\end{subequations}
where $\mathbf{E}_k$ denotes  the mean square error (MSE)  matrix for user $k$, defined as
\begin{align*} %\label{eq:E_k_corrected}
    &\mathbf{E}_k \triangleq  \left(\mathbf{I}_{\textcolor{black}{D_k}}-\mathbf{U}_k^H\mathbf{H}_k\boldsymbol{\Delta}\mathbf{H}^H\mathbf{X}_k\right) \left(\mathbf{I}_{\textcolor{black}{D_k}}-\mathbf{U}_k^H\mathbf{H}_k\boldsymbol{\Delta}\mathbf{H}^H\mathbf{X}_k\right)^H \nonumber \\
    & +\mathbf{U}_k^H\left(\sum_{j\neq k}\mathbf{H}_k\boldsymbol{\Delta}\mathbf{H}^H\mathbf{X}_j\mathbf{X}_j^H\left(\mathbf{H}_k\boldsymbol{\Delta}\mathbf{H}^H\right)^H + \mathbf{N}_k \right)\mathbf{U}_k
\end{align*}
The optimization problem  \eqref{eq:wmmse_reform_corrected} can be efficiently solved using a block coordinate descent (BCD) approach. The detailed update procedures are derived as follows:
\subsubsection{Update of Receiver Filters \(\{\mathbf{U}_k\}\)} For fixed $\{\mathbf{X}_k\}$ and \(\boldsymbol{\Delta}\), the optimal receive filter $\mathbf{U}_k$ that minimizes the MSE for user $k$ is given by:
    \begin{align}
    \label{U_cal1}
        \mathbf{U}_k = & \left(\sum_{j=1}^K \mathbf{H}_k\boldsymbol{\Delta}\mathbf{H}^H\mathbf{X}_j\mathbf{X}_j^H\left(\mathbf{H}_k\boldsymbol{\Delta}\mathbf{H}^H\right)^H + \mathbf{N}_k\right)^{-1} \nonumber \\
        & \times \mathbf{H}_k\boldsymbol{\Delta}\mathbf{H}^H\mathbf{X}_k.
    \end{align}
\subsubsection{Update of Weight Matrices \(\{\mathbf{W}_k\}\)} For fixed $\{\mathbf{U}_k\}$, $\{\mathbf{X}_k\}$ and \(\boldsymbol{\Delta}\), the optimal weight matrix $\mathbf{W}_k$ is given by the inverse of the resulting MSE matrix, i.e.,
    \begin{equation}
    \label{W_cal1}
        \mathbf{W}_k = \mathbf{E}_k^{-1} = \left(\mathbf{I}_{\textcolor{black}{D_k}} - \mathbf{U}_k^H\mathbf{H}_k\boldsymbol{\Delta}\mathbf{H}^H\mathbf{X}_k\right)^{-1}.
    \end{equation}

\subsubsection{Update of Precoding Coefficients \(\{\mathbf{X}_k\}\)} For fixed $\{\mathbf{U}_k\}$, $\{\mathbf{W}_k\}$, and \(\boldsymbol{\Delta}\), the subproblem for optimizing the low-dimensional precoding  coefficients $\{\mathbf{X}_k\}$ reduces to a convex quadratic program. By setting the gradient of the Lagrangian objective with respect to $\mathbf{X}_k$ to zero, the optimal solution is obtained as
    \begin{align}
    \label{X_cal1}
        \mathbf{X}_k = & \left(\sum_{i=1}^K \alpha_i \mathbf{H}\boldsymbol{\Delta}\mathbf{H}_i^H\mathbf{U}_i\mathbf{W}_i\mathbf{U}_i^H\mathbf{H}_i\boldsymbol{\Delta}\mathbf{H}^H + \mu \mathbf{H}\boldsymbol{\Delta}\mathbf{H}^H \right)^{-1} \nonumber \\
        & \times \alpha_k \mathbf{H}\boldsymbol{\Delta}\mathbf{H}_k^H\mathbf{U}_k\mathbf{W}_k,
    \end{align}
where the regularization parameter $ \mu \triangleq \sum_{i=1}^K \frac{\sigma_i^2}{P_{\max}} \mathrm{Tr}(\alpha_i\mathbf{U}_i\mathbf{W}_i\mathbf{U}_i^H)$ arises from the transmit power constraint.

\subsubsection{Update of Beam Selection Matrix \(\boldsymbol{\Delta}\)}
To optimize  $\boldsymbol{\Delta}$, we isolate the terms in the objective function \eqref{eq:wmmse_reform_corrected_a} that depend on $\boldsymbol{\Delta}$. This leads to the following subproblem:
    \begin{subequations}\label{eqn:f_delta_problem}
    \begin{align}
        \min_{\boldsymbol{\Delta}} \quad & f(\boldsymbol{\Delta})  \\
    \mathrm{s.t.} \quad & [\boldsymbol{\Delta}]_{ii} \in \{0,1\}, \quad \forall i, \\
    & \sum_{i=1}^M [\boldsymbol{\Delta}]_{ii} = K_s,
    \end{align}
    \end{subequations}
where the objective function $f(\boldsymbol{\Delta})$ is defined in \eqref{eq:f_delta_simplified_no_vars}.
    
\begin{figure*}[!t]
%\hrulefill
\begin{align} \label{eq:f_delta_simplified_no_vars}
f(\boldsymbol{\Delta}) \triangleq & \: \mathrm{Tr}\left( \left(\sum_{k=1}^K \alpha_k \mathbf{H}_k^H \mathbf{U}_k \mathbf{W}_k \mathbf{U}_k^H \mathbf{H}_k\right) \boldsymbol{\Delta} \left(\mathbf{H}^H \left(\sum_{j=1}^K \mathbf{X}_j\mathbf{X}_j^H\right) \mathbf{H}\right) \boldsymbol{\Delta}^H \right) \nonumber \\
& - 2\mathrm{Re}\left\{\mathrm{Tr}\left( \left(\sum_{k=1}^K \alpha_k \mathbf{H}^H\mathbf{X}_k \mathbf{W}_k\mathbf{U}_k^H\mathbf{H}_k\right) \boldsymbol{\Delta} \right)\right\} + \mu \mathrm{Tr}\left( \mathbf{H}^H \left(\sum_{i=1}^K \mathbf{X}_i\mathbf{X}_i^H\right) \mathbf{H} \boldsymbol{\Delta} \right)
\end{align}
\hrulefill
\vspace{-1.2em}
\end{figure*}

\begin{figure*}[!t]
%\hrulefill
\begin{align} \label{eq:grad_f}
\textcolor{black}{\operatorname{diag}\!\left(\nabla_{\boldsymbol{\Delta}}f(\boldsymbol{\Delta})\right)} = & \: 2 \mathrm{Re}\left\{ \mathrm{diag}\left( \left(\sum_{k=1}^K \alpha_k \mathbf{H}_k^H \mathbf{U}_k \mathbf{W}_k \mathbf{U}_k^H \mathbf{H}_k\right) \boldsymbol{\Delta} \left(\mathbf{H}^H \left(\sum_{j=1}^K \mathbf{X}_j \mathbf{X}_j^H\right) \mathbf{H}\right) \right) \right\} \nonumber \\
& - 2 \mathrm{Re}\left\{ \mathrm{diag}\left( \sum_{k=1}^K \alpha_k \mathbf{H}^H \mathbf{X}_k \mathbf{W}_k \mathbf{U}_k^H \mathbf{H}_k \right) \right\} + \mu \mathrm{diag}\left( \mathbf{H}^H \left(\sum_{j=1}^K \mathbf{X}_j \mathbf{X}_j^H\right) \mathbf{H} \right)
\end{align}
\hrulefill
\vspace{-1.5em}
\end{figure*}

The binary constraints on $\boldsymbol{\Delta}$ render problem \eqref{eqn:f_delta_problem} combinatorial and NP-hard. To address this difficulty, we adopt a penalty-based relaxation approach. Specifically, we first relax the binary constraints to the box constraints $0 \leq [\boldsymbol{\Delta}]_{ii}\leq 1$ for all $i$. To further promote binary solutions, we add the concave quadratic penalty term $-\beta \mathrm{Tr}(\boldsymbol{\Delta}^2)$ to the objective function, where $\beta > 0$ is a penalty parameter. The resulting relaxed penalized problem is given by
\begin{subequations}\label{eqn delta problem}
\begin{align}
\min_{\boldsymbol{\Delta}} \quad &  \mathcal{L}(\boldsymbol{\Delta}) \triangleq f(\boldsymbol{\Delta}) - \beta \mathrm{Tr}(\boldsymbol{\Delta}^2) \label{eqn delta problem 1} \\
\mathrm{s.t.} \quad & 0 \leq [\boldsymbol{\Delta}]_{ii} \leq 1, \quad \forall i, \label{eqn delta problem 3} \\
& \sum_{i=1}^{M}[\boldsymbol{\Delta}]_{ii} = K_{s}. \label{eqn delta problem 4}
\end{align}
\end{subequations}

\textcolor{black}{To analyze the exactness of the penalized relaxation, define}
$$
\begin{aligned}
\mathbf{\Omega}\triangleq \operatorname{Re}\!\Bigg\{&
\left(\sum_{k=1}^{K}\alpha_k \mathbf{H}_k^{H}\mathbf{U}_k\mathbf{W}_k\mathbf{U}_k^{H}\mathbf{H}_k\right) \\
&\odot
\left(\mathbf{H}^{H}\left(\sum_{j=1}^{K}\mathbf{X}_j\mathbf{X}_j^{H}\right)\mathbf{H}\right)^{T}
\Bigg\}.
\end{aligned}
$$
\textcolor{black}{The following proposition establishes a sufficient condition for the exactness of the penalized relaxation.}

{\color{black}
\begin{proposition}[Exactness of the Penalized Relaxation]
\label{prop:penalty_equivalence}
For fixed $\{\mathbf{U}_k\}$, $\{\mathbf{W}_k\}$, and $\{\mathbf{X}_k\}$, if
$\beta>\lambda_{\max}(\mathbf{\Omega})$, then every global minimizer of the relaxed penalized problem \eqref{eqn delta problem} is binary. Moreover, problems \eqref{eqn delta problem} and \eqref{eqn:f_delta_problem} have the same set of globally optimal selection matrices; their objective values differ only by the constant $-\beta K_s$ at such matrices. The proof is provided in Appendix B.
\end{proposition}
}

    To efficiently solve \eqref{eqn delta problem} , we employ the MM framework. At  iteration $t$,  a linear upper bound of the concave objective $\mathcal{L}(\boldsymbol{\Delta})$ is constructed around the current point $\boldsymbol{\Delta}^{(t)}$ using the first-order Taylor expansion:
    \begin{align}
    \mathcal{L}(\boldsymbol{\Delta}) \leq \mathcal{L}(\boldsymbol{\Delta}^{(t)}) + \mathrm{Tr}\left( (\nabla_{\boldsymbol{\Delta}}\mathcal{L}(\boldsymbol{\Delta}^{(t)}))^T (\boldsymbol{\Delta} - \boldsymbol{\Delta}^{(t)}) \right).
    \end{align}
    The gradient of the original objective $f(\boldsymbol{\Delta})$ with respect to the diagonal elements of $\boldsymbol{\Delta}$ is given in \eqref{eq:grad_f}.

Consequently, minimizing the linear upper bound is equivalent to solving the following LP:
    \begin{align}
    \min_{\boldsymbol{\Delta}} \quad & \mathrm{Re}\left\{\mathrm{Tr}\left( (\mathbf{C}^{(t)})^H \boldsymbol{\Delta} \right)\right\} \label{eqn:linearized_again} \\
    \mathrm{s.t.} \quad & \text{Constraints } \eqref{eqn delta problem 3}, \eqref{eqn delta problem 4}, \nonumber
    \end{align}
\textcolor{black}{where $\mathbf{C}^{(t)} \triangleq \nabla_{\boldsymbol{\Delta}}\mathcal{L}(\boldsymbol{\Delta}^{(t)}) = \left.\nabla_{\boldsymbol{\Delta}}f(\boldsymbol{\Delta})\right|_{\boldsymbol{\Delta}=\boldsymbol{\Delta}^{(t)}}-2\beta\boldsymbol{\Delta}^{(t)}$, and the required diagonal entries of $\nabla_{\boldsymbol{\Delta}}f(\boldsymbol{\Delta}^{(t)})$ are given by \eqref{eq:grad_f}.} Let $\mathbf{c}^{(t)} \triangleq \mathrm{Re}\{\mathrm{diag}(\mathbf{C}^{(t)})\}$ denote the vector of real-valued diagonal gradient elements, and define $\boldsymbol{\delta} \triangleq \mathrm{diag}(\boldsymbol{\Delta})$. The LP objective in \eqref{eqn:linearized_again} reduces to minimizing the inner product $(\mathbf{c}^{(t)})^T \boldsymbol{\delta}$. 
Accordingly, the optimal solution is obtained in closed form by selecting the indices corresponding to the $K_s$ smallest elements of $\mathbf{c}^{(t)}$, i.e., setting the corresponding diagonal entries of $\boldsymbol{\Delta}$ to 1, and setting all remaining entries to 0.

\textcolor{black}{To avoid reducing the number of active beams directly from $M$ to $K_s$ in a single iteration, the beam-selection update uses a current number of active beams $K_{\mathrm{cur}}$, which is initialized as $M$ and decreased by $K_{\mathrm{step}}$ at each outer iteration until it reaches $K_s$, after which the beam selection matrix is kept fixed. During progressive support reduction, $K_s$ in the cardinality constraint is temporarily replaced by the prescribed integer $K_{\mathrm{cur}}$ until the target support size is reached, and the exactness result in Proposition~\ref{prop:penalty_equivalence} remains applicable at each stage under its stated sufficient condition, with $K_s$ replaced by $K_{\mathrm{cur}}$.} The complete procedure of the proposed ALLSP algorithm is summarized in Algorithm \ref{alg:alg1}.

\begin{algorithm}[t]
\caption{Proposed ALLSP Algorithm}
\label{alg:alg1}
\begin{algorithmic}[1]
\STATE \textcolor{black}{\textbf{Initialize:} Iteration index $n=0$, max iterations $N_{\max}$. Initialize precoders $\{\mathbf{X}_k^{(0)}\}$, set $\boldsymbol{\Delta}^{(0)}=\mathbf{I}_M$ and $K_{\mathrm{cur}}=M$, and choose a positive integer support-reduction step $K_{\mathrm{step}}$.}

\REPEAT
    \STATE $n \leftarrow n+1$.
    \STATE {Update receiver filter $\mathbf{U}_k^{(n)}$ according to \eqref{U_cal1}.}
    
    \STATE {Update weight matrix $\mathbf{W}_k^{(n)}$ according to \eqref{W_cal1}.}
    
    \STATE {Update precoder $\mathbf{X}_k^{(n)}$ according to \eqref{X_cal1}.}

    \IF{\textcolor{black}{$K_{\mathrm{cur}}>K_s$}}
    \STATE \quad \textcolor{black}{$K_{\mathrm{cur}}\leftarrow\max\{K_s,K_{\mathrm{cur}}-K_{\mathrm{step}}\}$.}
    \STATE \quad {Update beam selection matrix $\boldsymbol{\Delta}^{(n)}$:}
    \STATE \quad \textcolor{black}{Evaluate \eqref{eq:grad_f} at $\boldsymbol{\Delta}^{(n-1)}$ and add $-2\beta\operatorname{diag}(\boldsymbol{\Delta}^{(n-1)})$ to obtain $\operatorname{diag}(\mathbf{C}^{(n-1)})$.}
    \STATE \quad \textcolor{black}{Identify the indices of the $K_{\mathrm{cur}}$ smallest elements in $\mathrm{Re}\{\operatorname{diag}(\mathbf{C}^{(n-1)})\}$.}
    \STATE \quad Set the corresponding diagonal elements of $\boldsymbol{\Delta}^{(n)}$ to 1, others to 0.
    \ELSE
    \STATE \quad \textcolor{black}{$\boldsymbol{\Delta}^{(n)}\leftarrow\boldsymbol{\Delta}^{(n-1)}$.}
    \ENDIF
    
    %\STATE \textbf{Update $\boldsymbol{\Delta}^{(n)}$ by setting the diagonal elements corresponding to the $K_s$ smallest entries of $\mathbf{c}^{(0)}$ to 1, and the rest to 0.}
\UNTIL{\textcolor{black}{($K_{\mathrm{cur}}=K_s$ and the overall objective function converges) or $n=N_{\max}$}}
\STATE \textbf{Output:} Optimized variables $\{\mathbf{U}_k^*\}$, $\{\mathbf{W}_k^*\}$, $\{\mathbf{X}_k^*\}$, and $\boldsymbol{\Delta}^*$.
\end{algorithmic}
\end{algorithm}

\section{Downlink Angle-User-Level Sparse Precoding Problem}
\subsection{Problem Formulation}
In the previous section, the beam selection matrix $\boldsymbol{\Delta}$ is shared among all users. To further enhance system flexibility and improve performance, we now propose an angle-user-level sparse precoding scheme in this section.

%\subsection{Problem Formulation}
%In the previous section, the beam selection matrix $\boldsymbol{\Delta}$ is shared among all users. To further enhance system flexibility and improve performance, we now consider an \textit{angle-user-level sparse precoding} scheme.
Specifically, each user $k$  is assigned an individual sparse diagonal selection matrix $\boldsymbol{\Delta}_k \in \mathbb{R}^{M \times M}$, allowing different users to activate different sets of angular beams. Denote the angle-domain precoder for user $k$  by $\mathbf{P}_k = \boldsymbol{\Delta}_k \mathbf{V}_k$. The received signal vector at user $k$ can then be expressed as
\begin{align}
\mathbf{y}_k = \mathbf{H}_k \boldsymbol{\Delta}_k \mathbf{V}_k \mathbf{s}_k + \sum_{j \neq k} \mathbf{H}_k \boldsymbol{\Delta}_j \mathbf{V}_j \mathbf{s}_j + \mathbf{n}_k.
\end{align}
Our objective is to maximize the WSR subject to per-user sparsity constraints and a total transmit power constraint. The resulting optimization problem is formulated as:
\begin{subequations}
\label{pro:angleuser_org}
\begin{align}
\!\!\!\!\max_{\{\mathbf{V}_k\},\{\boldsymbol{\Delta}_k\}} &  \sum_{k = 1}^K \alpha_k \log \det \Big( \mathbf{I}_{N_k} + \mathbf{H}_{k} \boldsymbol{\Delta}_k \mathbf{V}_k \mathbf{V}_k^H \boldsymbol{\Delta}_k^H \mathbf{H}_{k}^{H} \nonumber \\
& \!\!\times \Big(\sum_{j\ne k} \mathbf{H}_{k} \boldsymbol{\Delta}_j \mathbf{V}_j \mathbf{V}_j^H \boldsymbol{\Delta}_j^H \mathbf{H}_{k}^{H}+\sigma_{k}^{2} \mathbf{I}_{N_k} \Big)^{-1} \Big) \\
\mathrm{s.t.} \quad & \sum_{k=1}^K\mathrm{Tr} \left( \boldsymbol{\Delta}_k \mathbf{V}_k \mathbf{V}_k^H \boldsymbol{\Delta}_k^H \right) \le P_{\max}, \\
& [\boldsymbol{\Delta}_k]_{ii} \in \{0,1\}, \quad \forall i , \forall k, \\
& \sum_{i=1}^{M}[\boldsymbol{\Delta}_k]_{ii} = K_{s}, \quad \forall k.
\end{align}
\end{subequations}

\begin{remark}
\textbf{(Motivation and Theoretical Distinction)}
It is crucial to highlight two fundamental differences between this scheme and the angle-level approach discussed in Section III. 
First, regarding the {degrees of freedom}: Since our objective is to reduce the computational cost of \textit{signal weighting} (matrix-vector multiplications) rather than to physically deactivate hardware modules (e.g., switching off RF chains), the active beam set does not strictly need to be identical for all users. This distinction allows us to liberate the degrees of freedom in the spatial domain, enabling a flexible design where each user selects a unique subset of beams aligned with their specific channel characteristics.
Second, regarding the {solution structure}: The shift from a common selection matrix $\boldsymbol{\Delta}$ to user-specific matrices $\{\boldsymbol{\Delta}_k\}$ fundamentally alters the interference coupling. Consequently, the universal low-dimensional subspace structure derived in Proposition~\ref{prop:subspace} cannot be directly applied to this generalized case. This necessitates a rigorous re-analysis to establish whether a similar optimal subspace representation exists for angle-user-level sparsity.
\end{remark}

\subsection{Optimal Sparse Precoder Structure and Problem Reformulation}
In this subsection, we analyze the structure of the optimal precoding matrix for the angle-user-level WSR maximization problem in \eqref{pro:angleuser_org}. We demonstrate that the optimal solution lies in a specific low-dimensional subspace defined  by the channel matrices and the beam selection matrices.

\begin{proposition}[Subspace Characterization]
\label{prop:subspace_user}
The optimal angle-user-level sparse precoding matrix for user $k$, denoted by $\boldsymbol{\Delta}_k^*\mathbf{V}_k^*$, lies in the subspace spanned by the effective channel Hermitian $\boldsymbol{\Delta}_k^*\mathbf{H}^H$. Specifically, it satisfies the following structural relationship:
\begin{align}
\boldsymbol{\Delta}_k^*\mathbf{V}_k^* = \boldsymbol{\Delta}_k^*\mathbf{H}^H\mathbf{X}_k^*,
\label{eq:precoder_subspace_user}
\end{align}
where $\mathbf{H} \in \mathbb{C}^{N \times M}$ denotes the aggregate angle-domain channel matrix, and $\mathbf{X}_k \in \mathbb{C}^{N \times D_k}$ represents the low-dimensional projection coefficient matrix. The proof is provided in Appendix C.
\end{proposition}

\begin{remark}
\textbf{(Structural Comparison and Multi-Cell Extension)}
It is instructive to compare the result in \eqref{eq:precoder_subspace_user} with that of Proposition~\ref{prop:subspace}. While Proposition~\ref{prop:subspace} establishes a single, \textit{shared} low-dimensional subspace determined by the common active beam set $\boldsymbol{\Delta}$, Proposition~\ref{prop:subspace_user} reveals a more generalized structure where each user's optimal precoder resides in a \textit{user-specific} subspace spanned by the effective channel $\boldsymbol{\Delta}_k^*\mathbf{H}^H$. This indicates that the optimal signal direction for each user is strictly confined to the subspace formed by the interaction between the aggregate channel and their specific activated beams. 
Furthermore, it is worth noting that the theoretical conclusions derived in both Proposition~\ref{prop:subspace} and Proposition~\ref{prop:subspace_user} are not limited to single-cell scenarios. They can be readily extended to multi-cell cooperative sparse precoding\cite{hong2013joint,xu2023joint} to significantly reduce the optimization dimensionality. However, detailed formulations and derivations for the multi-cell scenario are omitted here due to space limitations.
\end{remark}

By substituting \eqref{eq:precoder_subspace_user} into  problem  \eqref{pro:angleuser_org} and absorbing the power constraint into the effective noise term, similar to Section III, we obtain the following equivalent optimization formulation for this problem:

% \begin{subequations}
% \label{eq:equivalent_problem2}
% \begin{align}
% \!\!\max_{\{\mathbf{X}_k\}, \{\boldsymbol{\Delta}_k\}}  & \sum_{k=1}^K \alpha_k \log \det \Big( \mathbf{I}_{N_k} + \mathbf{H}_k\boldsymbol{\Delta}_k\mathbf{H}^H\mathbf{X}_k\mathbf{X}_k^H\mathbf{H}\boldsymbol{\Delta}_k\mathbf{H}_k^H \nonumber\\
% & \!\!\!\!\!\!\!\!\!\!\!\!\!\!\!\!\!\!\!\!\!\times \Big( \sum_{j \neq k} \mathbf{H}_k\boldsymbol{\Delta}_j\mathbf{H}^H\mathbf{X}_j\mathbf{X}_j^H\mathbf{H}\boldsymbol{\Delta}_j\mathbf{H}_k^H + \sigma_k^2\mathbf{I}_{N_k} \Big)^{-1} \Big) \\
% \mathrm{s.t.} \quad & \sum_{k=1}^K \mathrm{Tr}\big(\mathbf{X}_k^H \mathbf{H}\boldsymbol{\Delta}_k\mathbf{H}^H \mathbf{X}_k\big) \leq P_{\max}, \label{eq:equivalent_problem2b}\\
% & [\boldsymbol{\Delta}_k]_{ii} \in \{0,1\},\ \forall i, \forall k, \\
% & \sum_{i=1}^M [\boldsymbol{\Delta}_k]_{ii} = K_s, \forall k.
% \end{align}
% \end{subequations}

%To decouple the sum-power constraint in \eqref{eq:equivalent_problem2b}, we adopt a  scale-invariant reformulation  similar to section III. By absorbing the power constraint into the effective noise term, problem \eqref{eq:equivalent_problem2} can be equivalently transformed into  the following unconstrained optimization problem:
\begin{subequations}
\label{eq:equivalent_problem1_unconstrained}
\begin{align}
 \!\!\max_{\{\mathbf{X}_k\}, \{\boldsymbol{\Delta}_k\}}  & \sum_{k=1}^K \alpha_k \log \det \Big( \mathbf{I}_{N_k} + \mathbf{H}_k\boldsymbol{\Delta}_k\mathbf{H}^H\mathbf{X}_k\mathbf{X}_k^H\mathbf{H}\boldsymbol{\Delta}_k\mathbf{H}_k^H \nonumber\\
& \!\!\!\!\!\!\!\!\!\!\!\!\!\!\!\!\!\times \Big( \sum_{j \neq k} \mathbf{H}_k\boldsymbol{\Delta}_j\mathbf{H}^H\mathbf{X}_j\mathbf{X}_j^H\mathbf{H}\boldsymbol{\Delta}_j\mathbf{H}_k^H + \mathbf{N}_k \Big)^{-1} \Big) \\
\mathrm{s.t.} \quad & [\boldsymbol{\Delta}_k]_{ii} \in \{0,1\},\ \forall i,\forall k, \\
& \sum_{i=1}^M [\boldsymbol{\Delta}_k]_{ii} = K_s, \forall k.
\end{align}
\end{subequations}
where the effective noise covariance matrix is defined as $\mathbf{N}_k \triangleq \frac{\sigma_k^2}{P_{\max}} \big( \sum_{j=1}^{K} \mathrm{Tr}( \mathbf{X}_j^H \mathbf{H}\boldsymbol{\Delta}_j\mathbf{H}^H \mathbf{X}_j ) \big) \mathbf{I}_{N_k}$. The optimal solution of problem \eqref{pro:angleuser_org} can be recovered from the optimal solution $(\{\mathbf{X}_k^*\}, \{\boldsymbol{\Delta}_k^*\})$ of problem \eqref{eq:equivalent_problem1_unconstrained} through the following scaling relationship:
\begin{align}
    \boldsymbol{\Delta}_k^*\mathbf{V}_k^* = \sqrt{\omega} \boldsymbol{\Delta}_k^*\mathbf{H}^H \mathbf{X}_k^*,
\end{align}
where the scaling factor $\omega$ is chosen to ensure that the  total transmit power constraint is satisfied with equality:
\begin{align}
    \omega \triangleq \frac{P_{\max}}{\sum_{j=1}^{K} \mathrm{Tr}\big( (\mathbf{X}_j^*)^H \mathbf{H}\boldsymbol{\Delta}_j^*\mathbf{H}^H \mathbf{X}_j^* \big)}.
\end{align}
Problem \eqref{eq:equivalent_problem1_unconstrained} is still a MINLP due to the binary user-specific selection matrices $\{\boldsymbol{\Delta}_k\}$. To solve it efficiently, we develop the AULLSP algorithm in the next subsection.

\subsection{Angle-User-Level Low-Complexity Sparse Precoding Algorithm Design}
To solve the \textcolor{black}{equivalent problem}  \eqref{eq:equivalent_problem1_unconstrained}, we extend the WMMSE  framework developed earlier. Specifically,  problem \eqref{eq:equivalent_problem1_unconstrained} can be equivalently reformulated as the following  weighted MSE minimization problem:
\begin{subequations}\label{eqn:wmmse_unconstrained}
\begin{align}
\min_{\substack{\{\mathbf{W}_k\},\{\mathbf{U}_k\}, \\ \{\mathbf{X}_k\},\{\boldsymbol{\Delta}_k\}}} \quad & \sum_{k=1}^K \alpha_k \left( \mathrm{Tr}\left(\mathbf{W}_k\mathbf{E}_k\right) - \log\det\left(\mathbf{W}_k\right) \right) \\
\mathrm{s.t.} \quad & [\boldsymbol{\Delta}_k]_{ii}\in\{0,1\},\quad \forall i, k, \\
& \sum_{i=1}^M[\boldsymbol{\Delta}_k]_{ii}=K_{s}, \quad \forall k,
\end{align}
\end{subequations}
where $\{\mathbf{U}_k\}$ and $\{\mathbf{W}_k\}$ are auxiliary variables. The MSE matrix for user $k$, denoted by $\mathbf{E}_k$, is defined as:
\begin{align}
\mathbf{E}_{k} &\triangleq  (\mathbf{I}_{\textcolor{black}{D_k}}-\mathbf{U}_k^H\mathbf{H}_k\boldsymbol{\Delta}_k \mathbf{H}^H\mathbf{X}_k)(\mathbf{I}_{\textcolor{black}{D_k}}-\mathbf{U}_k^H\mathbf{H}_k\boldsymbol{\Delta}_k \mathbf{H}^H\mathbf{X}_k)^H  \nonumber \\
& +\mathbf{U}_k^H\left(\sum_{j\neq k}\mathbf{H}_k\boldsymbol{\Delta}_j \mathbf{H}^H\mathbf{X}_j(\mathbf{H}_k\boldsymbol{\Delta}_j \mathbf{H}^H\mathbf{X}_j)^H + \mathbf{N}_k \right)\mathbf{U}_k.\nonumber
\end{align}
This problem can be solved using a BCD approach. Specifically, the updating rules for $\{\mathbf{U}_k\}$, $\{\mathbf{W}_k\}$, and $\{\mathbf{X}_k\}$ are analogous to those in the angle-level scheme discussed in Section III, with necessary adjustments to incorporate the user-specific beam selection matrices. The resulting iterative algorithm proceeds as follows:

\subsubsection{Update of Receiver Filters \(\{\mathbf{U}_k\}\)}
\begin{align}
\label{U_cal1user}
\mathbf{U}_k^* = &\left( \sum_{j=1}^K (\mathbf{H}_k \boldsymbol{\Delta}_j \mathbf{H}^H) \mathbf{X}_j \mathbf{X}_j^H (\mathbf{H}_k \boldsymbol{\Delta}_j \mathbf{H}^H)^H + \mathbf{N}_k \right)^{-1} \nonumber\\
&\times (\mathbf{H}_k \boldsymbol{\Delta}_k \mathbf{H}^H) \mathbf{X}_k.
\end{align}

\subsubsection{Update of Weight Matrices \(\{\mathbf{W}_k\}\)}
\begin{equation}
\label{W_cal1user}
\mathbf{W}_k^* = \mathbf{E}_k^{-1} = \left( \mathbf{I}_{\textcolor{black}{D_k}} - \mathbf{U}_k^H (\mathbf{H}_k \boldsymbol{\Delta}_k \mathbf{H}^H) \mathbf{X}_k \right)^{-1}.
\end{equation}

\subsubsection{Update of Precoding Coefficients \(\{\mathbf{X}_k\}\)}
\begin{align}
\label{X_cal1user}
\mathbf{X}_k = & \bigg( \sum_{i=1}^K \alpha_i (\mathbf{H}_i \boldsymbol{\Delta}_k \mathbf{H}^H)^H \mathbf{U}_i \mathbf{W}_i \mathbf{U}_i^H (\mathbf{H}_i \boldsymbol{\Delta}_k \mathbf{H}^H)  \nonumber \\
&  + \gamma \mathbf{H}\boldsymbol{\Delta}_k\mathbf{H}^H \bigg)^{-1}\times \alpha_k (\mathbf{H}_k \boldsymbol{\Delta}_k \mathbf{H}^H)^H \mathbf{U}_k \mathbf{W}_k,
\end{align}
where the regularization scalar $\gamma\triangleq \sum_{i=1}^K \frac{ \sigma_i^2}{P_{\max}} \mathrm{Tr}(\alpha_i\textcolor{black}{\mathbf{U}_i\mathbf{W}_i\mathbf{U}_i^H})$.

\subsubsection{Update of Beam Selection Matrices \{\(\boldsymbol{\Delta}_k\)\}}
For fixed $\{\mathbf{U}_k\}$, $\{\mathbf{W}_k\}$ and  $\{\mathbf{X}_k\}$, the optimization of the beam selection matrices $\{\boldsymbol{\Delta}_k\}_{k=1}^K$ decouples into $K$ independent subproblems. This decoupling arises because the global WMMSE objective function in \eqref{eqn:wmmse_unconstrained} can be rearranged into a sum of separable terms, where each term $f_k(\boldsymbol{\Delta}_k)$ depends solely on the selection matrix of user $k$. As a result, these subproblems can be solved in parallel.

Following the same relaxation and penalty strategy as in the angle-level sparse case, the subproblem associated with user $k$ can be written as
\begin{subequations}
\begin{align}
    \min_{\boldsymbol{\Delta}_k} \quad & \mathcal{L}_k(\boldsymbol{\Delta}_k) \triangleq f_k(\boldsymbol{\Delta}_k) - \beta \mathrm{Tr}(\boldsymbol{\Delta}_k^2) \label{eq:prob_f_k} \\
\mathrm{s.t.} \quad & 0 \leq [\boldsymbol{\Delta}_k]_{ii} \leq 1,\quad \forall i, k, \label{eq:prob_f_k_box} \\
& \sum_{i=1}^M[\boldsymbol{\Delta}_k]_{ii}=K_{s}, \quad \forall k, \label{eq:prob_f_k_cardinality}
\end{align}
\end{subequations}
where the binary constraints have been relaxed to continuous box constraints and a concave quadratic penalty term $-\beta \mathrm{Tr}(\boldsymbol{\Delta}_k^2)$ is introduced to promote binary solutions. The user-specific objective function $f_k(\boldsymbol{\Delta}_k)$ aggregates the signal distortion experienced by user $k$ and the interference caused by user $k$ to all other users, and is defined in \eqref{user_gg}. 
\begin{figure*}[!t]
\begin{align}
\label{user_gg}
f_k(\boldsymbol{\Delta}_k) \triangleq & \ \mathrm{Tr}\left(  \left(\sum_{i=1}^K \alpha_i \mathbf{H}_i^H \mathbf{U}_i \mathbf{W}_i \mathbf{U}_i^H \mathbf{H}_i \right)\boldsymbol{\Delta}_k \left( \mathbf{H}^H \mathbf{X}_k \mathbf{X}_k^H \mathbf{H} \right) \boldsymbol{\Delta}_k^H \right) \nonumber \\
& - 2 \mathrm{Re}\left\{\mathrm{Tr}\left( \boldsymbol{\Delta}_k \left( \alpha_k \mathbf{H}^H \mathbf{X}_k \mathbf{W}_k \mathbf{U}_k^H \mathbf{H}_k \right) \right)\right\} + \gamma \mathrm{Tr}\left( \boldsymbol{\Delta}_k \left( \mathbf{H}^H \mathbf{X}_k \mathbf{X}_k^H \mathbf{H} \right) \right).
\end{align}
\hrulefill
\vspace{-1.2em}
\end{figure*}

This resulting concave minimization problem is solved using the MM algorithm. At each inner iteration, the objective function is linearized around the current iterate $\boldsymbol{\Delta}_k^{(t)}$ using a first-order approximation, which yields the following LP:
%$\mathcal{L}_k(\boldsymbol{\Delta}_k)$ around the current iterate $\boldsymbol{\Delta}_k^{(t)}$, resulting in the following Linear Program (LP):
\begin{align}
    \min_{\boldsymbol{\Delta}_k} \quad & \mathrm{Re}\left\{\mathrm{Tr}\left((\mathbf{C}_k^{(t)})^H\boldsymbol{\Delta}_k\right)\right\} \label{eq:lp_delta_k} \\
    \mathrm{s.t.} \quad & \text{Constraints } \eqref{eq:prob_f_k_box}, \eqref{eq:prob_f_k_cardinality}. \nonumber
\end{align}
where $\mathbf{C}_k^{(t)}$ denotes the gradient of the penalized objective function evaluated at $\boldsymbol{\Delta}_k^{(t)}$:
\begin{align}\label{eq:penalized_gradient_user}
    \mathbf{C}_k^{(t)} = \nabla_{\boldsymbol{\Delta}_k}f_k(\boldsymbol{\Delta}_k)\big|_{\boldsymbol{\Delta}_k=\boldsymbol{\Delta}_k^{(t)}} - 2\beta\boldsymbol{\Delta}_k^{(t)}.
\end{align}
The optimal solution to \eqref{eq:lp_delta_k} is obtained in closed form by selecting the indices corresponding to the $K_s$ smallest real-valued diagonal elements of $\mathbf{C}_k^{(t)}$ and setting the corresponding diagonal entries to 1, while setting all remaining entries to zero.%The gradient $\nabla_{\boldsymbol{\Delta}_k}f_k(\boldsymbol{\Delta}_k)$ is given in \eqref{eq:grad_g}.

\begin{figure*}[!t]
\begin{align} \label{eq:grad_g}
\textcolor{black}{\operatorname{diag}\!\left(\nabla_{\boldsymbol{\Delta}_k} f_k(\boldsymbol{\Delta}_k)\right)} = & \: 2 \mathrm{Re}\left\{ \mathrm{diag}\left( \left(\sum_{i=1}^K \alpha_i \mathbf{H}_i^H \mathbf{U}_i \mathbf{W}_i \mathbf{U}_i^H \mathbf{H}_i \right) \boldsymbol{\Delta}_k \left(\mathbf{H}^H \mathbf{X}_k \mathbf{X}_k^H \mathbf{H}\right) \right) \right\} \nonumber \\
& - 2 \mathrm{Re}\left\{ \mathrm{diag}\left( \alpha_k (\mathbf{H}^H\mathbf{X}_k) (\mathbf{W}_k\mathbf{U}_k^H\mathbf{H}_k) \right) \right\} + \gamma \cdot \mathrm{diag}\left( \mathbf{H}^H \mathbf{X}_k\mathbf{X}_k^H\mathbf{H} \right)
\end{align}
\hrulefill
\vspace{-1.5em}
\end{figure*}

\textcolor{black}{Following the same progressive support-reduction strategy as in Algorithm~\ref{alg:alg1}, $K_s$ in the user-specific cardinality constraints is temporarily replaced by $K_{\mathrm{cur}}$ until the target support size is reached.} The complete procedure of the proposed AULLSP algorithm is summarized in Algorithm \ref{alg:alg2}.

\begin{algorithm}[t]
\caption{Proposed AULLSP Algorithm}
\label{alg:alg2}
\begin{algorithmic}[1]
\STATE \textcolor{black}{\textbf{Initialize:} Iteration index $n=0$, max iterations $N_{max}$. Initialize precoders $\{\mathbf{X}_k^{(0)}\}$ and set $\boldsymbol{\Delta}_{k}^{(0)}=\mathbf{I}_M$, $\forall k$. Set $K_{\mathrm{cur}}=M$ and choose a positive integer support-reduction step $K_{\mathrm{step}}$.}
\REPEAT
\STATE $n \leftarrow n+1$.
\STATE Update receiver filters  $\mathbf{U}_k^{(n)}$ according to \eqref{U_cal1user}.
\STATE Update weight matrices $\mathbf{W}_k^{(n)}$ according to \eqref{W_cal1user}.
\STATE Update precoders $\mathbf{X}_k^{(n)}$ according to \eqref{X_cal1user}.
\IF{\textcolor{black}{$K_{\mathrm{cur}}>K_s$}}
    \STATE \quad \textcolor{black}{$K_{\mathrm{cur}}\leftarrow\max\{K_s,K_{\mathrm{cur}}-K_{\mathrm{step}}\}$.}
    \STATE \quad {Update beam selection matrices $\{\boldsymbol{\Delta}_k^{(n)}\}$:}
    \STATE \quad \textcolor{black}{Calculate $\operatorname{diag}(\mathbf{C}_k^{(n-1)})$ according to \eqref{eq:penalized_gradient_user}, with the required gradient diagonal given by \eqref{eq:grad_g}.}
    \STATE \quad \textcolor{black}{Identify the indices of the $K_{\mathrm{cur}}$ smallest elements in $\mathrm{Re}\{\operatorname{diag}(\mathbf{C}_k^{(n-1)})\}$.}
    \STATE \quad Set the corresponding diagonal elements of $\boldsymbol{\Delta}_k^{(n)}$ to 1, others to 0.
\ELSE
    \STATE \quad \textcolor{black}{$\boldsymbol{\Delta}_k^{(n)}\leftarrow\boldsymbol{\Delta}_k^{(n-1)}$, $\forall k$.}
\ENDIF
\UNTIL{\textcolor{black}{($K_{\mathrm{cur}}=K_s$ and the overall objective function converges) or $n=N_{\max}$}}
\STATE \textbf{Output:} Optimized variables $\{\mathbf{U}_k^*\}$, $\{\mathbf{W}_k^*\}$, $\{\mathbf{X}_k^*\}$,  and $\{\boldsymbol{\Delta}_{k}^{*}\}$.
\end{algorithmic}
\end{algorithm}

\section{Complexity Analysis}
\label{sec:complexity}

We compare the considered methods from two perspectives: the computational complexity of precoder design and the signal weighting cost over one channel coherence interval.

%\subsection{Precoder Design Complexity}
\textcolor{black}{Regarding precoder-design complexity, the proposed ALLSP and AULLSP exploit the low-dimensional subspace structures established in Propositions~\ref{prop:subspace} and~\ref{prop:subspace_user}, respectively, reducing the required matrix inversion to dimension $N\times N$. Their per-iteration computational cost comprises $\mathcal{O}(N^3)$ for the matrix inversion and $\mathcal{O}(MN^2)$ for channel-dependent matrix products. Since $M\gg N$ in massive MIMO systems, the overall per-iteration complexity is dominated by $\mathcal{O}(MN^2)$. By contrast, the conventional sparsity-constrained WMMSE precoding method in~\cite{hong2013joint}, referred to as S-WMMSE, operates in the full-dimensional precoder space and requires the inversion of an $M\times M$ matrix at each iteration, resulting in a per-iteration complexity of $\mathcal{O}(M^3)$. Its iterative sparsity-promoting updates introduce further computational overhead.}

%\subsection{Signal Weighting Cost}
The signal weighting cost is the real-time complexity of applying the precoder to the data vector over a coherence interval. For a dense fully-digital precoder, the multiplication $\mathbf{x}=\mathbf{P}_{\mathrm{ant}}\mathbf{s}$ requires $M D$ complex multiplications per symbol, yielding a total cost of $N_{sym} M D$ over one coherence interval, where $N_{sym}$ is the total number of resource elements requiring precoding. For the proposed sparse angle-domain implementation, signal weighting consists of two steps: sparse angle-domain multiplication and transformation back to the antenna domain via IFFT. With $K_s$ active beams, the resulting total cost is
\begin{equation}
N_{sym}\left(K_s D + \frac{M}{2}\log_2 M\right).
\label{eq:signal_weighting_cost}
\end{equation}
Therefore, the proposed sparse architecture substantially reduces the per-symbol weighting cost, and this saving is amplified by the large factor $N_{sym}$.

\textcolor{black}{To quantify this benefit, consider $M=128$, $D=16$, $K_s=48$, and a support-reduction step of $K_{\mathrm{step}}=10$ under the 8-RB setting described in Section~II-B. Dense and sparse signal weighting require $2{,}048$ and $1{,}216$ complex multiplications per resource element, respectively, corresponding to a $40.625\%$ reduction. Multiplying the per-resource-element saving of 832 by $N_{sym}$ gives cumulative signal-weighting savings of approximately $9.58\times10^6$, $3.83\times10^7$, and $1.53\times10^8$ complex multiplications over precoder-update intervals of 5, 20, and 80~ms, respectively. In AULLSP, the common low-rank factors used to evaluate the user-specific gradients in \eqref{eq:grad_g} are computed once per support update and reused across users, while the remaining products are evaluated using each user's $D_k$-column precoder block $\mathbf{X}_k$. Using a multiplication-only convention in which four real multiplications are counted as one equivalent complex multiplication, the gradient-diagonal evaluations in \eqref{eq:grad_f} and \eqref{eq:grad_g}, together with the associated covariance updates, require approximately $1.23\times10^5$ and $1.31\times10^5$ equivalent complex multiplications per support update for ALLSP and AULLSP, respectively. Over the $(M-K_s)/K_{\mathrm{step}}=8$ support updates, the larger cumulative overhead is approximately $1.05\times10^6$ equivalent complex multiplications per precoder computation. Conservatively deducting this overhead once per interval, the net savings for the three intervals remain approximately $8.53\times10^6$, $3.73\times10^7$, and $1.52\times10^8$ equivalent complex multiplications, respectively.}

\section{NUMERICAL RESULTS}
In this section, we present numerical simulation results to evaluate the performance of the proposed ALLSP and AULLSP schemes. Unless otherwise specified, we consider a  MU-MIMO downlink system in which the BS is equipped with $M = 128$ transmit antennas and the total number of transmitted data streams is  $D=16$. Each user terminal employs four receive antennas,  supporting the  transmission of four spatial data streams, i.e., $N_k = D_k = 4,\, \forall k$. The wireless propagation environment follows the urban macro (UMa) scenario. Specifically, the channel realizations are generated according to the ``3GPP TR 38.901 UMa-NLOS'' model using the QuaDRiGa toolbox~\cite{jaeckel2014quadriga}. The carrier frequency is set to 1~GHz, and the subcarrier spacing is configured to 30~kHz. \textcolor{black}{Users are independently and uniformly distributed within a $120^\circ$ sector, with their radial distances from the BS ranging from $0.1$ to $0.3$~km. The total transmit power is normalized to $P_{\max}=1$.} The user weights are set to $\alpha_k = 1,\ \forall k$. \textcolor{black}{For comparison, S-WMMSE is used as the conventional sparsity-constrained WMMSE benchmark, while the full-dimensional WMMSE precoder is adopted as a performance benchmark for the considered sparse precoding schemes.} \textcolor{black}{In addition, sparse beamspace precoding (SBP)~\cite{gonultacs2021hardware} and energy-maximum beam selection (EMBS)~\cite{tong2023low} are included as additional low-complexity sparse precoding baselines. The suffix ``-Ant'' denotes the corresponding antenna-domain implementation.} \textcolor{black}{Unless otherwise specified, ALLSP and AULLSP use fixed support-reduction steps of $K_{\mathrm{step}}=10$ and $8$ for $K_s=48$ and $64$, respectively, completing support reduction in eight outer iterations.} \textcolor{black}{The noise power is set to be equal for all users and, for each channel realization, is calculated as $\sigma_k^2=10^{\frac{1}{K}\sum_{i=1}^{K}\log_{10}\!\left(\frac{1}{N_i}\left\|\mathbf{H}_i\right\|_F^2\right)}\times 10^{-\frac{\mathrm{SNR}}{10}},\ \forall k$, where SNR denotes the average receive SNR in the dB domain over all users.}

\begin{figure}[t!]
    \centering
    \includegraphics[width=0.97\columnwidth]{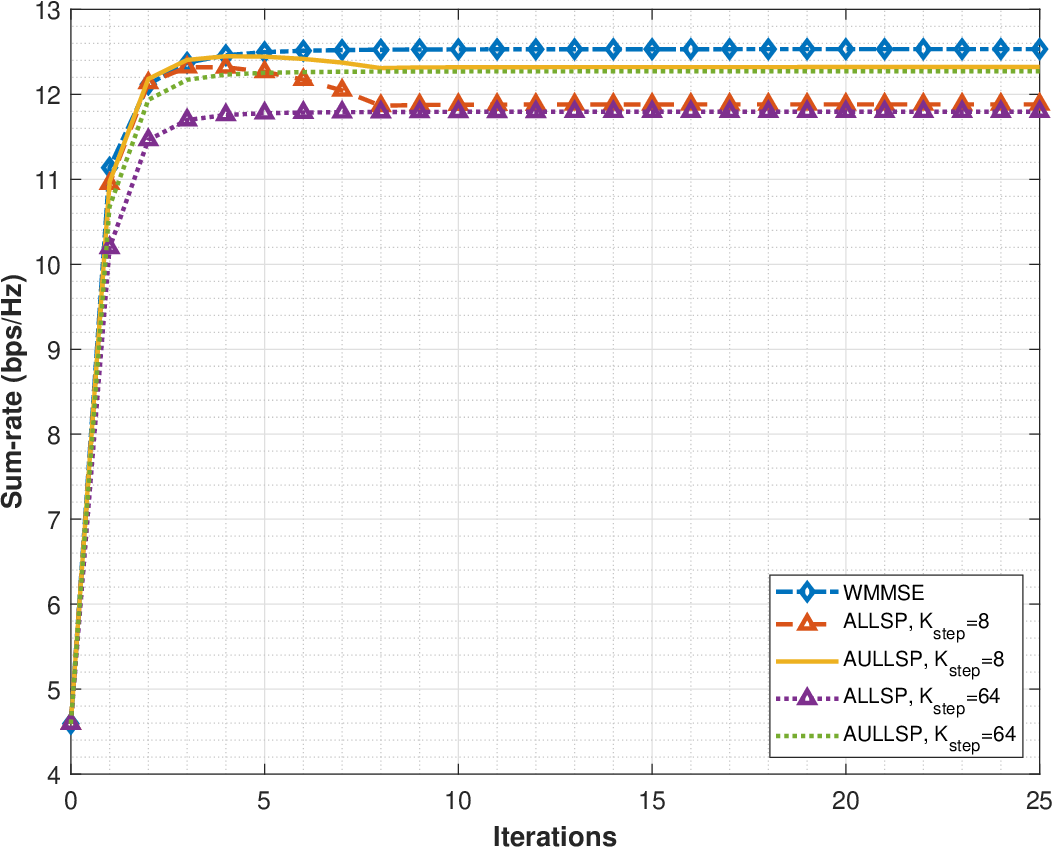}
    \caption{\textcolor{black}{Convergence behavior of the proposed ALLSP and AULLSP schemes.}}
    \label{fig:convergen}
\end{figure}

\textcolor{black}{Fig. \ref{fig:convergen} compares the convergence behavior of ALLSP and AULLSP under progressive and direct support reduction for $K_s=64$ at SNR $=10$ dB. All curves start from the same full-beam ZF initialization. The step size $K_{\mathrm{step}}=8$ reduces the working support from 128 to 64 over the first eight outer iterations, whereas $K_{\mathrm{step}}=64$ imposes the target support in the first iteration. The direct-reduction variants stabilize earlier, while the progressive variants become stable within approximately 10 iterations, with their initial variations caused by the changing support. At steady state, progressive reduction improves the sum-rate of ALLSP and AULLSP by approximately 0.73\% and 0.40\%, respectively, compared with direct reduction. AULLSP consistently outperforms ALLSP. Although progressive support reduction provides only a modest sum-rate improvement, its additional beam-selection overhead is small relative to the signal-weighting cost. We therefore adopt this strategy in the subsequent simulations. For each support size, the average number of beams removed per update is $(M-K_s)/8$. If this value is noninteger, it is rounded up for some updates and down for the others so that exactly $M-K_s$ beams are removed over eight updates.}

\begin{figure}[t!]
    \centering
    \subfigure[48 beams/antennas]{
        \label{fig:snr24}
        \includegraphics[width=0.97\columnwidth]{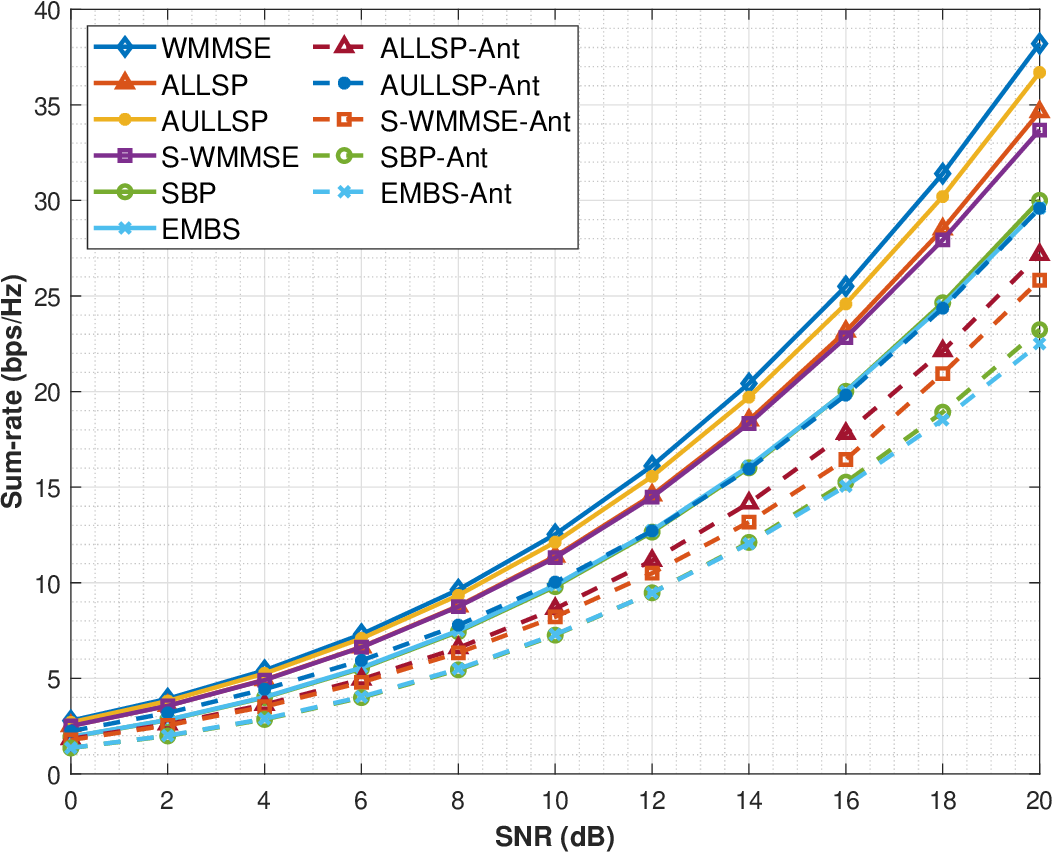}
    }
    \\[-0.5em]
    \subfigure[64 beams/antennas]{
        \label{fig:snr16}
        \includegraphics[width=0.97\columnwidth]{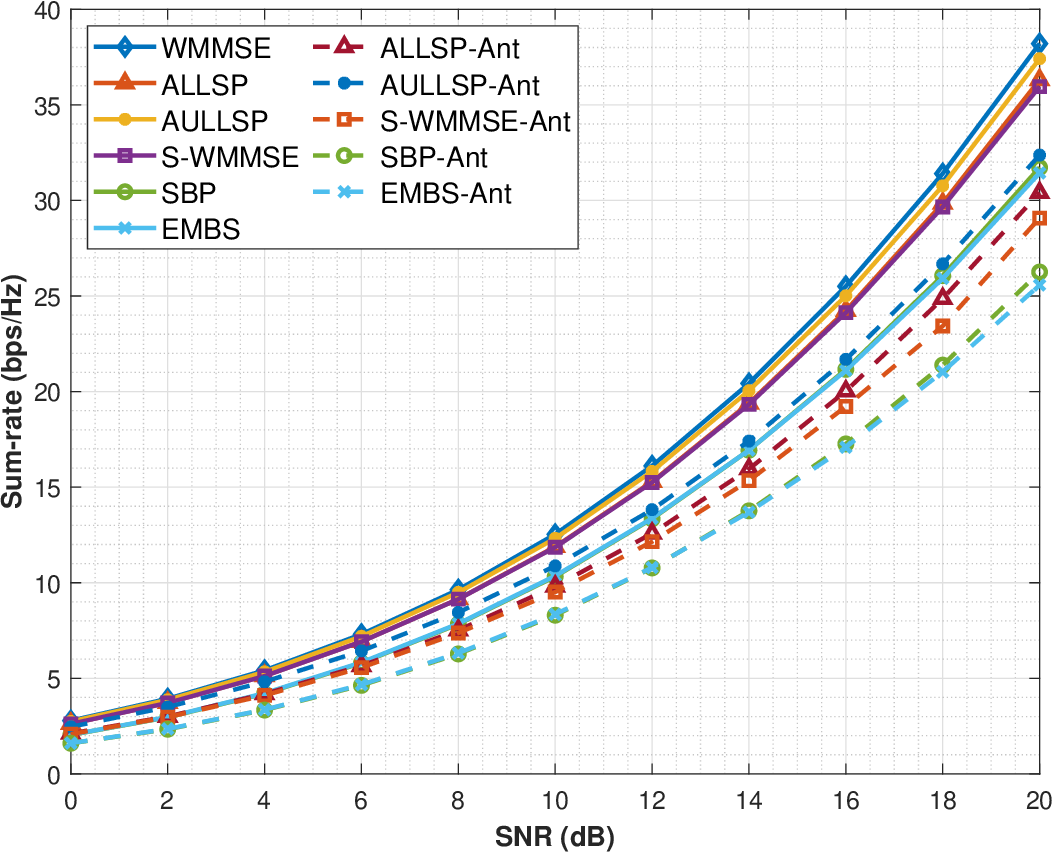}
    }
    \caption{Achievable sum-rate versus SNR under different sparsity levels.}
    \label{fig:snr}
    \vspace{-15pt}
\end{figure}

\textcolor{black}{Fig. \ref{fig:snr} shows the system sum-rate versus SNR under two sparsity levels, namely 62.5\% and 50\%, corresponding to 48 and 64 active beams, respectively.} AULLSP consistently outperforms ALLSP across both sparsity levels, confirming the effectiveness of the angle-user-level sparsity design. The proposed angle-domain sparse precoding schemes outperform their antenna-domain counterparts, highlighting the benefit of operating in the angle domain. Furthermore, both AULLSP and ALLSP achieve sum-rates close to that of WMMSE, showing that high spectral efficiency can still be maintained under sparse precoding. \textcolor{black}{For example, under the 50\% sparsity level, the sum-rate losses of AULLSP and ALLSP relative to WMMSE at 16 dB are only 1.94\% and 4.98\%, respectively. Across both sparsity levels, AULLSP consistently outperforms S-WMMSE, SBP, and EMBS, whereas ALLSP achieves performance comparable to S-WMMSE while clearly outperforming SBP and EMBS.} \textcolor{black}{According to the signal-weighting complexity expression in \eqref{eq:signal_weighting_cost}, the two sparsity settings considered here reduce the signal-weighting complexity by 40.625\% and 28.125\%, respectively, compared with dense precoding.} Overall, the results demonstrate that AULLSP and ALLSP maintain strong sum-rate performance under different sparsity levels while achieving substantial savings in signal weighting cost.

\begin{figure}[t!]
    \centering
    \subfigure[SNR $=10$ dB]{
        \label{fig:sparsity_scan_10}
        \includegraphics[width=0.97\columnwidth]{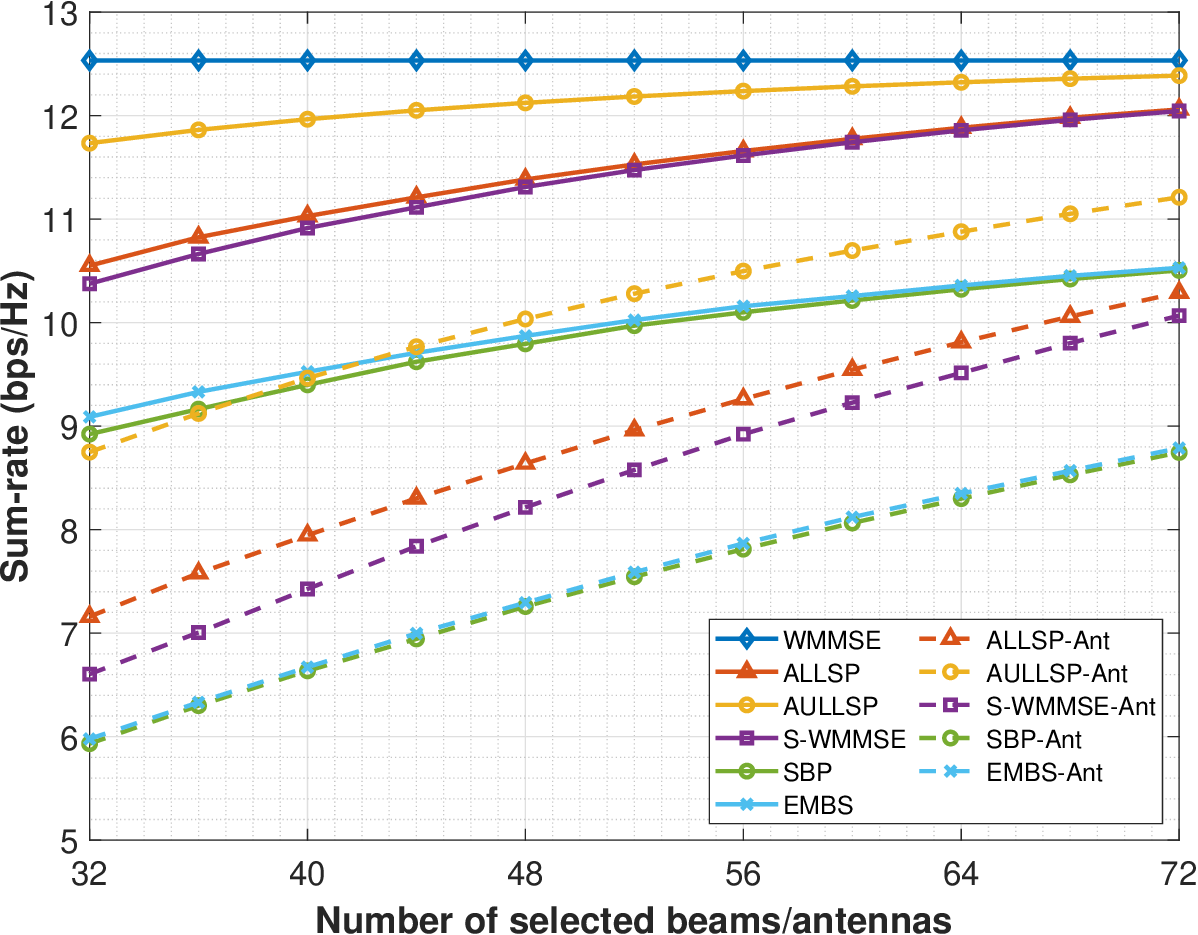}
    }
    \\[-0.5em]
    \subfigure[SNR $=20$ dB]{
        \label{fig:sparsity_scan_20}
        \includegraphics[width=0.97\columnwidth]{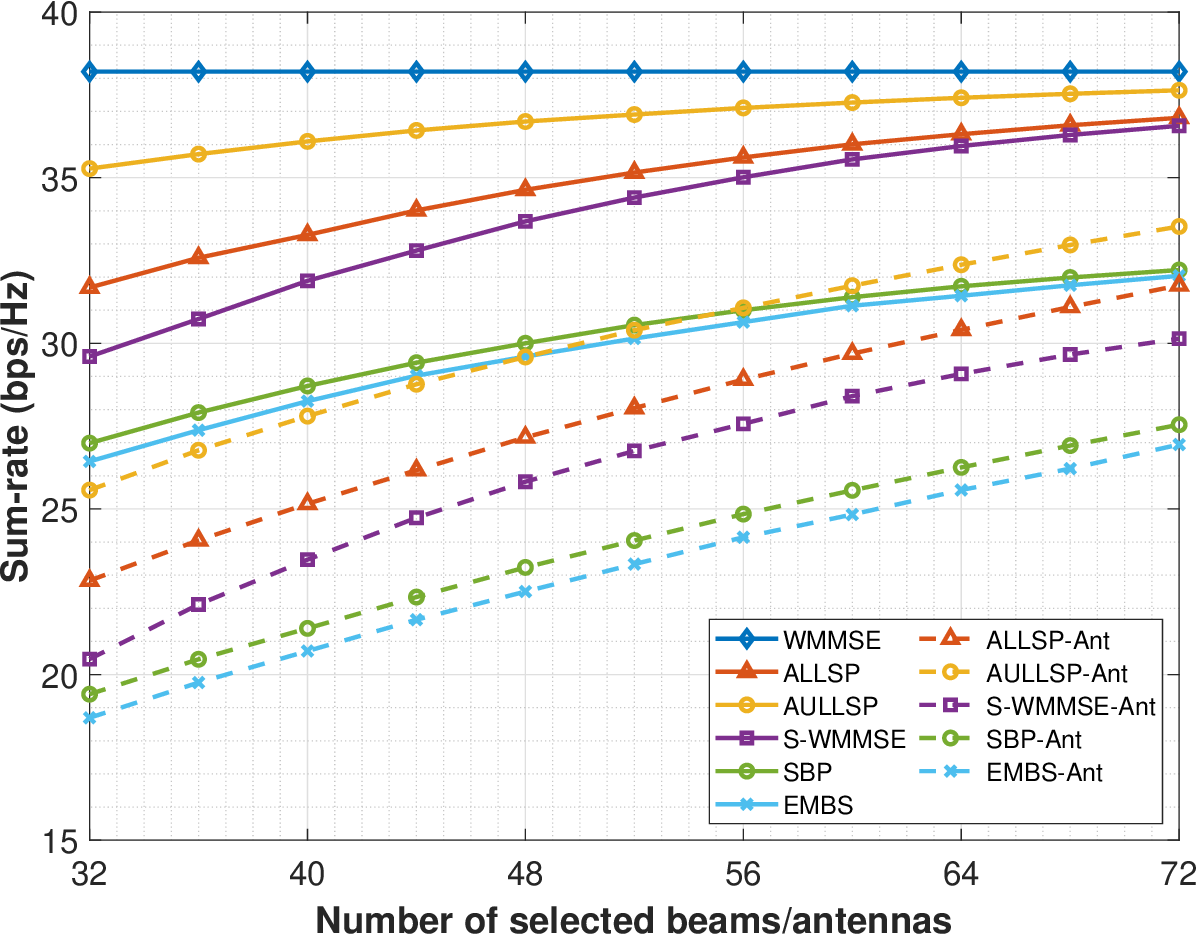}
    }
    \caption{Achievable sum-rate versus the number of selected beams/antennas at different SNRs.}
    \label{fig:sparsity_scan}
\end{figure}

\begin{figure}[tbp]
    \centering
    \includegraphics[width=0.97\columnwidth]{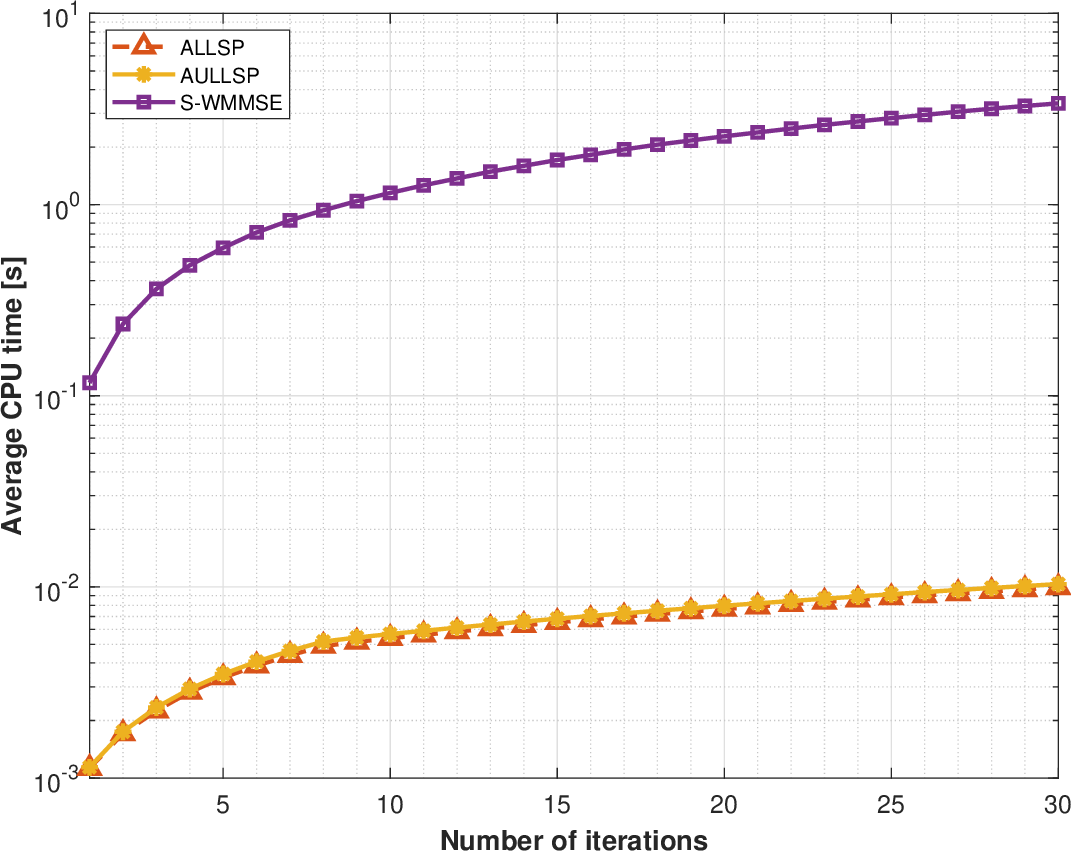}
    \caption{Average cumulative CPU time versus the number of iterations for three sparse WMMSE algorithms.}
    \label{fig:weightscope_combined}
\end{figure}

\textcolor{black}{Fig. \ref{fig:sparsity_scan} examines the performance--complexity tradeoff by varying the number of selected beams or antennas from 32 to 72, corresponding to sparsity levels from $75\%$ to $43.75\%$ for $M=128$. At both SNR $=10$ and $20$ dB, the sum-rate increases with the support size. AULLSP achieves the highest sum-rate among the sparse schemes and remains closest to the full-dimensional WMMSE benchmark. ALLSP performs below AULLSP, particularly at smaller support sizes, because its common support provides less design freedom than the user-specific supports of AULLSP. The angle-domain implementations consistently outperform their antenna-domain counterparts, while SBP and EMBS remain below the proposed schemes. EMBS performs slightly better at 10 dB, whereas SBP has a slight advantage at 20 dB. At 64 selected beams, the AULLSP losses relative to WMMSE are only $1.68\%$ and $2.07\%$ at 10 and 20 dB, respectively. Meanwhile, the signal-weighting cost reduction decreases from $53.125\%$ to $21.875\%$ as the support size increases from 32 to 72.}

\textcolor{black}{Fig. \ref{fig:weightscope_combined} compares the average cumulative CPU time of ALLSP, AULLSP, and S-WMMSE for $M=128$, $D=16$, $K_s=48$, and SNR $=20$ dB. The measurements were conducted in MATLAB R2022a on a 64-bit Windows 11 computer equipped with a 13th Gen Intel Core i9-13900H processor and 16 GB of RAM. After 10 outer iterations, ALLSP and AULLSP require only $5.39$ and $5.67$ ms, respectively, whereas S-WMMSE requires $1.15$ s. After 20 iterations, the corresponding times are $7.66$ ms, $7.97$ ms, and $2.28$ s. The large gap arises because ALLSP and AULLSP perform the WMMSE updates in the low-dimensional subspaces established in Propositions~\ref{prop:subspace} and~\ref{prop:subspace_user}, whereas S-WMMSE repeatedly solves high-dimensional systems through multiple inner reweighted updates.}

\section{Conclusion}
This paper proposes a baseband-efficient WMMSE framework for sparse precoding in fully-digital massive MU-MIMO systems. Unlike conventional WMMSE designs that mainly focus on precoder computation, the proposed framework explicitly accounts for the signal weighting cost in baseband processing under the sum-rate maximization criterion. Two sparse precoding models, namely angle-level sparsity and angle-user-level sparsity, are developed, where the latter further enlarges the design freedom through user-specific beam selection. For both models, we show that the optimal sparse precoder lies in a low-dimensional subspace determined by the channel structure and the corresponding selection matrices. \textcolor{black}{Based on this result, the original sparse WSR maximization problems are reformulated into reduced-dimensional forms.} \textcolor{black}{Compared with conventional sparsity-constrained WMMSE precoding, the resulting ALLSP and AULLSP algorithms substantially reduce the precoder-design complexity.} \textcolor{black}{Simulation results show that the proposed methods achieve sum-rate performance close to that of the full-dimensional WMMSE benchmark while reducing the signal-weighting cost relative to dense precoding.}

\section{appendix}
\begin{appendices}
\section{Proof of Proposition~\ref{prop:subspace}}
\label{appendix:prop1_proof}

We prove this proposition by contradiction. Let $\{\bm{\Delta}^*, \mathbf{V}_k^*\}$ be an optimal solution to problem (8). The objective function and the power constraint depend solely on the effective precoding matrices $\mathbf{P}_k^* = \bm{\Delta}^*\mathbf{V}_k^*$. The proposition states that each $\mathbf{P}_k^*$ must lie in the subspace $\mathcal{S} = \operatorname{span}(\bm{\Delta}^*\mathbf{H}^H)$.

Assume, for the sake of contradiction, that this statement does not hold. Then there exists at least one user index $k$ for which the optimal effective precoder $\mathbf{P}_k^*$ does not lie entirely in the subspace $\mathcal{S}$. Consequently, $\mathbf{P}_k^*$ can  be uniquely decomposed  into two orthogonal components:
\begin{align}
    \mathbf{P}_k^* = \mathbf{A}_k + \mathbf{B}_k,
\end{align}
where $\mathbf{A}_k$ is the orthogonal projection of $\mathbf{P}_k^*$ onto $\mathcal{S}$, and $\mathbf{B}_k$ is the projection onto its orthogonal complement $\mathcal{S}^{\perp}$. By  assumption, $\mathbf{B}_k \neq \mathbf{0}$ for at least one user $k$.

By definition, $\mathbf{P}_k^* = \bm{\Delta}^*\mathbf{V}_k^*$, which implies that $\mathbf{P}_k^*$ has zero rows corresponding to the zero entries of $\bm{\Delta}^*$. Equivalently, this yields $\mathbf{P}_k^* = \bm{\Delta}^*\mathbf{P}_k^*$. The basis vectors of the subspace $\mathcal{S}$ are given by the columns of $\bm{\Delta}^*\mathbf{H}^H$, all of which  share the same row-sparse structure induced by $\bm{\Delta}^*$. Consequently, any vector in $\mathcal{S}$, including the projection $\mathbf{A}_k$, must also \textcolor{black}{satisfy} this structure, i.e., $\mathbf{A}_k = \bm{\Delta}^*\mathbf{A}_k$. It  then follows that the orthogonal component $\mathbf{B}_k = \mathbf{P}_k^* - \mathbf{A}_k$ must also satisfy:
\begin{align}
    \bm{\Delta}^*\mathbf{B}_k &= \bm{\Delta}^*(\mathbf{P}_k^* - \mathbf{A}_k) = \bm{\Delta}^*\mathbf{P}_k^* - \bm{\Delta}^*\mathbf{A}_k \nonumber\\
    &= \mathbf{P}_k^* - \mathbf{A}_k = \mathbf{B}_k.
\end{align}
By the definition of orthogonality, $\mathbf{B}_k$ is orthogonal to every vector in the subspace $\mathcal{S}$. This implies
\begin{align}
    (\bm{\Delta}^*\mathbf{H}^H)^H \mathbf{B}_k = \mathbf{H}(\bm{\Delta}^*)^H \mathbf{B}_k = \mathbf{H}\bm{\Delta}^*\mathbf{B}_k = \mathbf{0}.
\end{align}
Substituting the relation $\mathbf{B}_k = \bm{\Delta}^*\mathbf{B}_k$ into above equation yields $\mathbf{H}\mathbf{B}_k = \mathbf{0}$, which implies  that $\mathbf{H}_j\mathbf{B}_k = \mathbf{0}$ for all users $j$. This result  shows that the orthogonal component $\mathbf{B}_k$ lies in the null space of  every user's channel and therefore contributes neither to the desired signal nor to inter-user interference. Consequently, $\mathbf{B}_k$ has no effect on the objective function:
\begin{align}
    \mathbf{H}_j \mathbf{P}_k^* = \mathbf{H}_j (\mathbf{A}_k + \mathbf{B}_k) = \mathbf{H}_j \mathbf{A}_k.
\end{align}
Thus, a solution constructed solely from the projected components $\{\mathbf{A}_k\}$ achieves exactly the  same sum-rate as the original solution.

We now examine the impact on the transmit power constraint. The total transmit  power of the original optimal solution is given by:
\begin{align}
    P_{\text{total}}^* &= \sum_{k=1}^{K} \operatorname{Tr}(\mathbf{P}_k^* (\mathbf{P}_k^*)^H) \nonumber \\
    &=  \sum_{k=1}^{K} \operatorname{Tr}\big((\mathbf{A}_k + \mathbf{B}_k)(\mathbf{A}_k + \mathbf{B}_k)^H\big) \nonumber \\
    &= \sum_{k=1}^{K} \operatorname{Tr}(\mathbf{A}_k (\mathbf{A}_k)^H) + \sum_{k=1}^{K} \operatorname{Tr}(\mathbf{B}_k (\mathbf{B}_k)^H) \le P_{\text{max}}.
\end{align}
Since we assumed that $\mathbf{B}_k \neq \mathbf{0}$ for at least one user $k$, it follows that $\operatorname{Tr}(\mathbf{B}_k (\mathbf{B}_k)^H) = \|\mathbf{B}_k\|_F^2 > 0$. As a result,  the total transmit power satisfies the following strict inequality:
\begin{align}
    \sum_{k=1}^{K} \operatorname{Tr}(\mathbf{A}_k (\mathbf{A}_k)^H) < P_{\text{max}}.
\end{align}
We can construct a new feasible solution that strictly improves upon the assumed optimal one. Define $\mathbf{P}_k' = \eta \mathbf{A}_k$, where the scaling factor $\eta > 1$ is chosen such that the transmit  power constraint is satisfied with equality, i.e., $\eta^2 \sum_{k=1}^{K} \operatorname{Tr}(\mathbf{A}_k (\mathbf{A}_k)^H) = P_{\text{max}}$. The resulting set of effective precoders $\{\mathbf{P}_k'\}$ is feasible for the original problem.  Specifically,  we can define the corresponding precoding matrices $\mathbf{V}_k' = \eta \mathbf{A}_k$. This construction is valid because, as shown earlier, $\mathbf{A}_k = \bm{\Delta}^*\mathbf{A}_k$, which implies $\mathbf{P}_k' = \eta \mathbf{A}_k = \eta \bm{\Delta}^*\mathbf{A}_k = \bm{\Delta}^*(\eta \mathbf{A}_k) = \bm{\Delta}^*\mathbf{V}_k'$.

The objective value achieved by this new feasible solution is given by:
\begin{align}
    &\sum_{k=1}^{K} \alpha_k \log\det\Big(\mathbf{I} + \mathbf{H}_k \mathbf{P}_k' (\mathbf{P}_k')^H \mathbf{H}_k^H  \nonumber \\ &\ \ \ \ \ \ \ \ \ \ \ \ \ \ \ \ \ \ \ \ \  \Big(\sum_{j \neq k} \mathbf{H}_k \mathbf{P}_j' (\mathbf{P}_j')^H \mathbf{H}_k^H + \sigma_k^2\mathbf{I}\Big)^{-1}\Big) \nonumber \\
    =& \sum_{k=1}^{K} \alpha_k \log\det\Big(\mathbf{I} + \mathbf{H}_k \mathbf{A}_k (\mathbf{A}_k)^H \mathbf{H}_k^H \nonumber \\ &\ \ \ \ \ \ \ \ \ \ \ \ \ \ \ \ \ \ \ \ \  \Big(\sum_{j \neq k} \mathbf{H}_k \mathbf{A}_j (\mathbf{A}_j)^H \mathbf{H}_k^H + \frac{\sigma_k^2}{\eta^2}\mathbf{I}\Big)^{-1}\Big).
\end{align}
Since $\eta > 1$, the effective noise term $\sigma_k^2/\eta^2$ is strictly smaller than the original noise variance $\sigma_k^2$. Consequently, the newly constructed solution achieves a strictly larger weighted sum-rate than that obtained using the precoders based on $\{\mathbf{A}_k\}$, which themselves yield the same weighted sum-rate as the presumed optimal solution $\{\mathbf{P}_k^*\}$.

This result contradicts the assumption that $\{\bm{\Delta}^*, \mathbf{V}_k^*\}$ constitutes an optimal solution. Therefore, the initial assumption must be false, and the orthogonal component $\mathbf{B}_k$ must be zero for all $k$. It follows that the optimal effective precoder $\mathbf{P}_k^* = \bm{\Delta}^*\mathbf{V}_k^*$ must lie entirely in the subspace spanned by $\bm{\Delta}^*\mathbf{H}^H$, which completes the proof.

\vspace{-5pt}
{\color{black}
\section{Proof of Proposition~\ref{prop:penalty_equivalence}}
\label{appendix:penalty_equivalence}

We prove Proposition~\ref{prop:penalty_equivalence} by first expressing the
penalized objective as a quadratic function of the diagonal entries of
$\boldsymbol{\Delta}$ and then characterizing its global minimizers over
the relaxed feasible set. Throughout the proof, $\{\mathbf{U}_k\}$,
$\{\mathbf{W}_k\}$, and $\{\mathbf{X}_k\}$ are held fixed. To isolate the
quadratic dependence on $\boldsymbol{\Delta}$, we introduce
\begin{subequations}
\begin{align}
\mathbf{\Phi} &\triangleq \sum_{k=1}^{K}\alpha_k
\mathbf{H}_k^{H}\mathbf{U}_k\mathbf{W}_k\mathbf{U}_k^{H}\mathbf{H}_k, \\
\mathbf{\Psi} &\triangleq \mathbf{H}^{H}\left(\sum_{j=1}^{K}
\mathbf{X}_j\mathbf{X}_j^{H}\right)\mathbf{H}.
\end{align}
\end{subequations}
Both $\mathbf{\Phi}$ and $\mathbf{\Psi}$ are Hermitian positive
semidefinite matrices. Recall that $\boldsymbol{\Delta}$ is real and
diagonal. Hence, by letting
$\boldsymbol{\delta}=\operatorname{diag}(\boldsymbol{\Delta})$, we have
$\boldsymbol{\Delta}=\operatorname{Diag}(\boldsymbol{\delta})$ and
$\boldsymbol{\Delta}^{H}=\boldsymbol{\Delta}$. We now use the identity
\begin{align}
&\operatorname{Tr}\!\left(\mathbf{M}\operatorname{Diag}(\boldsymbol{\delta})
\mathbf{N}\operatorname{Diag}(\boldsymbol{\delta})\right) \nonumber\\
&\hspace{2.5em}=\boldsymbol{\delta}^{T}
(\mathbf{M}\odot\mathbf{N}^{T})\boldsymbol{\delta},
\label{eq:diag_trace_identity}
\end{align}
which holds for any conformable matrices $\mathbf{M}$ and $\mathbf{N}$.
Applying \eqref{eq:diag_trace_identity} to the first term of
\eqref{eq:f_delta_simplified_no_vars} yields
\begin{align}
\operatorname{Tr}(\mathbf{\Phi}\boldsymbol{\Delta}
\mathbf{\Psi}\boldsymbol{\Delta})
=\boldsymbol{\delta}^{T}\mathbf{\Omega}\boldsymbol{\delta},
\label{eq:delta_quadratic_form}
\end{align}
where $\mathbf{\Omega}=\operatorname{Re}\{\mathbf{\Phi}\odot
\mathbf{\Psi}^{T}\}$ agrees with the definition preceding
Proposition~\ref{prop:penalty_equivalence}. Since $\mathbf{\Phi}$ and
$\mathbf{\Psi}$ are Hermitian, $\mathbf{\Omega}$ is a real symmetric
matrix, and the quadratic representation in
\eqref{eq:delta_quadratic_form} is therefore real-valued.

The remaining terms of $f(\boldsymbol{\Delta})$ are affine in
$\boldsymbol{\delta}$. Collecting their coefficients gives the real vector
\begin{align}
\mathbf{q}\triangleq{}&-2\operatorname{Re}\!\left\{
\operatorname{diag}\!\left(\sum_{k=1}^{K}\alpha_k
\mathbf{H}^{H}\mathbf{X}_k\mathbf{W}_k\mathbf{U}_k^{H}\mathbf{H}_k
\right)\right\} \nonumber\\
&+\mu\operatorname{diag}(\mathbf{\Psi}).
\label{eq:delta_linear_coefficient}
\end{align}
Combining this linear term with \eqref{eq:delta_quadratic_form}, the
original and penalized objectives can be written, respectively, as
\begin{subequations}
\begin{align}
f(\boldsymbol{\Delta})
&=\boldsymbol{\delta}^{T}\mathbf{\Omega}\boldsymbol{\delta}
+\mathbf{q}^{T}\boldsymbol{\delta}, \\
\mathcal{L}(\boldsymbol{\Delta})
&=\boldsymbol{\delta}^{T}(\mathbf{\Omega}-\beta\mathbf{I})
\boldsymbol{\delta}+\mathbf{q}^{T}\boldsymbol{\delta}.
\label{eq:penalized_delta_quadratic}
\end{align}
\end{subequations}
It follows directly from \eqref{eq:penalized_delta_quadratic} that the
Hessian of $\mathcal{L}$ with respect to the real diagonal variable is
\begin{align}
\nabla_{\boldsymbol{\delta}}^{2}\mathcal{L}
=2(\mathbf{\Omega}-\beta\mathbf{I}).
\label{eq:penalty_hessian}
\end{align}
When $\beta>\lambda_{\max}(\mathbf{\Omega})$, every eigenvalue of
$\mathbf{\Omega}-\beta\mathbf{I}$ is strictly negative. Therefore,
\eqref{eq:penalty_hessian} is negative definite, and
$\mathcal{L}(\boldsymbol{\Delta})$ is strictly concave over the feasible
polytope
\begin{align}
\mathcal{D}\triangleq\{\boldsymbol{\delta}\in\mathbb{R}^{M}:\,
\mathbf{0}\preceq\boldsymbol{\delta}\preceq\mathbf{1},\;
\mathbf{1}^{T}\boldsymbol{\delta}=K_s\}.
\label{eq:delta_polytope}
\end{align}
where $K_s\in\{1,\ldots,M\}$ is a prescribed integer.
We next examine the geometry of $\mathcal{D}$. Its vertices are exactly
the binary vectors containing $K_s$ ones. Suppose, by contradiction, that
a global minimizer of \eqref{eqn delta problem} is not a vertex. It can
then be expressed as a strict convex combination of two distinct points
in $\mathcal{D}$. Strict concavity implies that the objective value at
this combination is strictly larger than the corresponding convex
combination of the two objective values. Since neither of those values
can be smaller than the global minimum, a contradiction is obtained.
Consequently, every global minimizer of \eqref{eqn delta problem} must be
a binary vertex.

It remains to compare the objective values of the two problems over these
binary vertices. For any binary feasible $\boldsymbol{\Delta}$, we have
\begin{align}
\operatorname{Tr}(\boldsymbol{\Delta}^{2})
=\sum_{m=1}^{M}\delta_m^{2}
=\sum_{m=1}^{M}\delta_m
=\operatorname{Tr}(\boldsymbol{\Delta})=K_s.
\label{eq:binary_penalty_constant}
\end{align}
Thus, the penalty term takes the same value $-\beta K_s$ at every binary
feasible point and does not alter their ordering according to
$f(\boldsymbol{\Delta})$. Together with the fact that all global
minimizers of the relaxed penalized problem are binary, this shows that
\eqref{eqn delta problem} and \eqref{eqn:f_delta_problem} have the same
globally optimal selection matrices, which completes the proof.
}

\vspace{-5pt}
\section{Proof of Proposition~\ref{prop:subspace_user}}
\label{appendix:prop_user_proof}

The proof also proceeds by contradiction, but the detailed construction differs from that of Proposition~\ref{prop:subspace} because the user-specific selection matrices $\{\bm{\Delta}_k^*\}_{k=1}^K$ induce different user-specific subspaces. We therefore only highlight the key differences here.

Let $\{\bm{\Delta}_k^*, \mathbf{V}_k^*\}_{k=1}^K$ be an optimal solution to problem \eqref{pro:angleuser_org}, and define the effective precoders as $\mathbf{P}_k^*=\bm{\Delta}_k^*\mathbf{V}_k^*$. Proposition~\ref{prop:subspace_user} claims that, for each user $k$, $\mathbf{P}_k^*$ lies in the subspace
\begin{equation}
\mathcal{S}_k = \operatorname{span}(\bm{\Delta}_k^*\mathbf{H}^H).
\end{equation}

Assume by contradiction that this is not true. Then there exists at least one user, say user $m$, such that $\mathbf{P}_m^* \notin \mathcal{S}_m$. We can decompose
\begin{equation}
\mathbf{P}_m^*=\mathbf{A}_m+\mathbf{B}_m,
\end{equation}
where $\mathbf{A}_m$ is the orthogonal projection of $\mathbf{P}_m^*$ onto $\mathcal{S}_m$, and $\mathbf{B}_m\neq \mathbf{0}$ lies in $\mathcal{S}_m^\perp$.

Since $\mathbf{P}_m^*=\bm{\Delta}_m^*\mathbf{V}_m^*$, we have $\mathbf{P}_m^*=\bm{\Delta}_m^*\mathbf{P}_m^*$. Moreover, because every vector in $\mathcal{S}_m$ has the row support induced by $\bm{\Delta}_m^*$, it follows that $\mathbf{A}_m=\bm{\Delta}_m^*\mathbf{A}_m$, and hence also $\mathbf{B}_m=\bm{\Delta}_m^*\mathbf{B}_m$. By orthogonality,
\begin{align}
    (\bm{\Delta}_m^*\mathbf{H}^H)^H \mathbf{B}_m = \mathbf{H}(\bm{\Delta}_m^*)^H \mathbf{B}_m = \mathbf{H}\bm{\Delta}_m^*\mathbf{B}_m = \mathbf{0}.
\end{align}
which, together with $\mathbf{B}_m=\bm{\Delta}_m^*\mathbf{B}_m$, implies
$\mathbf{H}\mathbf{B}_m=\mathbf{0}$. Therefore, $\mathbf{H}_j\mathbf{B}_m=\mathbf{0}$ for all users $j$, meaning that $\mathbf{B}_m$ contributes neither to the desired signal nor to the interference terms in the objective function.

Now define a new set of effective precoders by keeping all users unchanged except user $m$, namely,
\begin{subequations}
\begin{align}
\mathbf{P}_k' &= \mathbf{P}_k^*, \quad k\neq m,\\
\mathbf{P}_m' &= \mathbf{A}_m.
\end{align}
\end{subequations}
Since $\mathbf{H}_j\mathbf{B}_m=\mathbf{0}$ for all $j$, this new solution achieves exactly the same weighted sum-rate as the original one. On the other hand, because $\mathbf{B}_m\neq \mathbf{0}$, removing $\mathbf{B}_m$ strictly reduces the total transmit power. Hence, all effective precoders $\{\mathbf{P}_k'\}$ can be further scaled by a common factor $\eta>1$ so that the power constraint is met with equality, which strictly increases the weighted sum-rate and leads to a contradiction.

The scaled solution remains feasible under the same selection matrices $\{\bm{\Delta}_k^*\}$ because each $\mathbf{P}_k'$ preserves the row support induced by $\bm{\Delta}_k^*$. Therefore, the assumption is false, and $\mathbf{B}_m=\mathbf{0}$. Since the choice of user $m$ is arbitrary, we conclude that, for every user $k$, the optimal effective precoder $\mathbf{P}_k^*=\bm{\Delta}_k^*\mathbf{V}_k^*$ must lie entirely in the subspace spanned by $\bm{\Delta}_k^*\mathbf{H}^H$.
\end{appendices}

%\vspace{-1.5em}
\bibliographystyle{IEEEtran}
%\small\bibliography{ref}
\bibliography{ref}

\vfill

\end{document}